\documentclass[preprint,12pt]{emulateapj}

\usepackage{natbib}
\usepackage[T1]{fontenc}
\usepackage{graphicx}
\usepackage[space]{grffile}
\usepackage{latexsym}
\usepackage{amsfonts,amsmath,amssymb}
\usepackage{xstring}

\newcommand{\kepmag}{\textit{Kp}\xspace}

\usepackage{xspace}
\usepackage{amsmath}

\newcommand{\Kepler}{\textit{Kepler}\xspace} 
\newcommand{\ktwo}{\textit{K2}\xspace}
\newcommand{\TERRA}{\texttt{TERRA}\xspace}
\newcommand{\SM}{\texttt{SM}\xspace}
\newcommand{\iso}{\texttt{iso}\xspace}
\newcommand{\isochrones}{\texttt{isochrones}\xspace}

\newcommand{\Mstar}{\ensuremath{M_{\star}}\xspace}
\newcommand{\Rstar}{\ensuremath{R_{\star}}\xspace} 
\newcommand{\Lstar}{\ensuremath{L_{\star}}\xspace} 
\newcommand{\fe}{$[$Fe/H$]$\xspace}
\newcommand{\teff}{$T_{\mathrm{eff}}$\xspace}  
\newcommand{\logg}{\ensuremath{\log g}\xspace} 
\newcommand{\vsini}{\ensuremath{v \sin i}\xspace} 
\newcommand{\logrhk}{\ensuremath{\log R'_{\mbox{\scriptsize HK}}}\xspace} 

\newcommand{\Mp}{\ensuremath{M_{P}}\xspace} 
\newcommand{\Rp}{\ensuremath{R_P}\xspace}
\newcommand{\teq}{$T_{\mathrm{eq}}$\xspace}
\newcommand{\Sinc}{\ensuremath{S_{inc}}\xspace}

\newcommand{\ms}{m s$^{-1}$\xspace}
\newcommand{\kms}{m s$^{-1}$\xspace}
\newcommand{\msyr}{m s$^{-1}$ yr$^{-1}$\xspace}
\newcommand{\Se}{\ensuremath{S_{\oplus}}\xspace}
\newcommand{\Me}{\ensuremath{M_{\oplus}}\xspace} 
\renewcommand{\Re}{\ensuremath{R_{\oplus}}\xspace} 

\newcommand{\gcc}{g~cm$^{-3}$\xspace}
\newcommand{\Rsun}{\ensuremath{R_{\odot}}\xspace }
\newcommand{\Msun}{\ensuremath{M_{\odot}}\xspace}
\newcommand{\Lsun}{\ensuremath{L_{\odot}}\xspace} 
\newcommand{\rhostar}{\ensuremath{\rho_\star}\xspace}
\newcommand{\rhostarcirc}{\ensuremath{\rho_{\star,\mathrm{circ}}}\xspace}

\newcommand{\lonperi}{\ensuremath{\omega_{\star}}\xspace}
\newcommand{\ecosw}{\ensuremath{e \cos \lonperi}\xspace}
\newcommand{\esinw}{\ensuremath{e \sin \lonperi}\xspace}
\newcommand{\sqrtecosw}{\ensuremath{\sqrt{e} \cos \lonperi}\xspace}
\newcommand{\sqrtesinw}{\ensuremath{\sqrt{e} \sin \lonperi}\xspace}
\newcommand{\dvdt}{\ensuremath{dv/dt}\xspace}
\newcommand{\sigjit}{\ensuremath{\sigma_\mathrm{jit}}\xspace}

\newcommand{\loglike}{\ensuremath{\ln \mathcal{L}}\xspace}

\newcommand{\SpecMatch}{\texttt{SpecMatch}\xspace}

\newcommand{\stellar}[1]{%
    \IfEqCase{#1}{%
        {name}{K2-24\xspace}%
        {NAME}{K2-24\xspace}%
        {epic}{203771098\xspace}%
        {teff}{$5743 \pm 60$\xspace}%
        {logg}{$4.29 \pm 0.07$\xspace}%
        {fe}{$0.42 \pm 0.04$\xspace}%
        {Mstar}{$1.12 \pm 0.05$\xspace}%
        {Rstar}{$1.21 \pm 0.11$\xspace}%
        {nobsiodine}{32\xspace}%
        {rho}{$0.89 \pm 0.23$}%
        {spectype}{G3}%
    }[\PackageError{tree}{Undefined option to tree: #1}{}]%
}%
\newcommand{\Mcoreb}{\ensuremath{17.6\pm4.3}\xspace}
\newcommand{\Mcorec}{\ensuremath{16.1\pm4.2}\xspace}
\newcommand{\fenvb}{\ensuremath{24\pm8}\xspace}
\newcommand{\fenvc}{\ensuremath{48\pm9}\xspace}

\newcommand{\Mcorewaterb}{\ensuremath{18.0\pm4.9}\xspace}
\newcommand{\Mcorewaterc}{\ensuremath{17.2\pm5.0}\xspace}
\newcommand{\fenvwaterb}{\ensuremath{14\pm5\%}\xspace}
\newcommand{\fenvwaterc}{\ensuremath{36\pm8\%}\xspace}

\newcommand{\planet}[2]{%
\IfEqCase{#1}{
{c}{\IfEqCase{#2}{
{Rstar_on_a}{$0.019_{ -0.000 }^{ +0.001 }$}%
    {a}{$0.247\pm0.004$}%
    {b}{$0.22_{ -0.16 }^{ +0.17 }$}%
    {Rp}{$7.82\pm0.72$}%
    {rhostar}{$1.47_{ -0.23 }^{ +0.31 }$}%
    {esinom}{$0.00\pm0.00$}%
    {i}{$89.76_{ -0.21 }^{ +0.18 }$}%
    {k}{$4.6\pm1.2$}%
    {density}{$0.31\pm0.12$}%
    {ecosom}{$0.00\pm0.00$}%
    {t0}{$2082.6251\pm0.0004$}%
    {Sinc}{$24\pm5$}%
    {T14}{$6.47_{ -0.03 }^{ +0.04 }$}%
    {P}{$42.3633\pm0.0006$}%
    {mass}{$27.0\pm6.9$}%
    {teq}{$606\pm139$}%
    {Rp_on_Rstar}{$5.94_{ -0.04 }^{ +0.10 }$}%
    {T23}{$5.70_{ -0.06 }^{ +0.03 }$}%
    {e}{< 0.00 (95\%)}%
    }[\PackageError{tree}{Undefined option to tree: #2}{}]}%
    {b}{\IfEqCase{#2}{
{Rstar_on_a}{$0.035_{ -0.002 }^{ +0.005 }$}%
    {a}{$0.154\pm0.002$}%
    {b}{$0.37_{ -0.24 }^{ +0.22 }$}%
    {Rp}{$5.68\pm0.56$}%
    {rhostar}{$1.00_{ -0.33 }^{ +0.21 }$}%
    {esinom}{$0.00\pm0.00$}%
    {i}{$89.25_{ -0.61 }^{ +0.49 }$}%
    {k}{$4.5\pm1.1$}%
    {density}{$0.63\pm0.25$}%
    {ecosom}{$0.00\pm0.00$}%
    {t0}{$2072.7948\pm0.0007$}%
    {Sinc}{$60\pm14$}%
    {T14}{$5.48_{ -0.04 }^{ +0.07 }$}%
    {P}{$20.8851\pm0.0003$}%
    {mass}{$21.0\pm5.4$}%
    {teq}{$767\pm177$}%
    {Rp_on_Rstar}{$4.31_{ -0.08 }^{ +0.17 }$}%
    {T23}{$4.95_{ -0.11 }^{ +0.05 }$}%
    {e}{$0.00_{-0.00}^{+0.00}$}%
    }[\PackageError{tree}{Undefined option to tree: #2}{}]}%
    }[\PackageError{tree}{Undefined option to tree: #1}{}]%
}%

\newcommand{\planetecc}[2]{%
\IfEqCase{#1}{
{c}{\IfEqCase{#2}{
{Rstar_on_a}{$0.019_{ -0.000 }^{ +0.001 }$}%
    {a}{$0.247\pm0.004$}%
    {b}{$0.22_{ -0.16 }^{ +0.17 }$}%
    {Rp}{$7.82\pm0.72$}%
    {rhostar}{$1.47_{ -0.23 }^{ +0.31 }$}%
    {esinom}{$-0.02\pm0.15$}%
    {i}{$89.76_{ -0.21 }^{ +0.18 }$}%
    {k}{$5.3\pm1.1$}%
    {density}{$0.36\pm0.12$}%
    {ecosom}{$0.00\pm0.09$}%
    {t0}{$2082.6251\pm0.0004$}%
    {Sinc}{$24\pm5$}%
    {T14}{$6.47_{ -0.03 }^{ +0.04 }$}%
    {P}{$42.3633\pm0.0006$}%
    {mass}{$31.0\pm6.4$}%
    {teq}{$606\pm139$}%
    {Rp_on_Rstar}{$5.94_{ -0.04 }^{ +0.10 }$}%
    {T23}{$5.70_{ -0.06 }^{ +0.03 }$}%
    {e}{< 0.39 (95\%)}%
    }[\PackageError{tree}{Undefined option to tree: #2}{}]}%
    {b}{\IfEqCase{#2}{
{Rstar_on_a}{$0.035_{ -0.002 }^{ +0.005 }$}%
    {a}{$0.154\pm0.002$}%
    {b}{$0.37_{ -0.24 }^{ +0.22 }$}%
    {Rp}{$5.68\pm0.56$}%
    {rhostar}{$1.00_{ -0.33 }^{ +0.21 }$}%
    {esinom}{$-0.06\pm0.16$}%
    {i}{$89.25_{ -0.61 }^{ +0.49 }$}%
    {k}{$5.1\pm1.2$}%
    {density}{$0.70\pm0.26$}%
    {ecosom}{$0.20\pm0.09$}%
    {t0}{$2072.7948\pm0.0007$}%
    {Sinc}{$60\pm14$}%
    {T14}{$5.48_{ -0.04 }^{ +0.07 }$}%
    {P}{$20.8851\pm0.0003$}%
    {mass}{$23.2\pm5.3$}%
    {teq}{$767\pm177$}%
    {Rp_on_Rstar}{$4.31_{ -0.08 }^{ +0.17 }$}%
    {T23}{$4.95_{ -0.11 }^{ +0.05 }$}%
    {e}{$0.24_{-0.11}^{+0.11}$}%
    }[\PackageError{tree}{Undefined option to tree: #2}{}]}%
    }[\PackageError{tree}{Undefined option to tree: #1}{}]%
}%

\begin{document}

\title{Two Transiting Low Density Sub-Saturns from K2}
\shorttitle{Two Transiting Low Density Sub-Saturns from K2}

\author{
Erik A.\ Petigura\altaffilmark{1,10},
Andrew W.\ Howard\altaffilmark{2},
Eric D.\ Lopez\altaffilmark{3},
Katherine M.\ Deck\altaffilmark{1,11},
Benjamin J.\ Fulton\altaffilmark{2,12},
Ian J.\ M.\ Crossfield\altaffilmark{4,13},
David R.\ Ciardi\altaffilmark{5},
Eugene Chiang\altaffilmark{6,7},
Eve J.\ Lee\altaffilmark{6},
Howard Isaacson\altaffilmark{6},
Charles A.\ Beichman\altaffilmark{1},
Brad M.\ S.\ Hansen\altaffilmark{8},
Joshua E.\ Schlieder\altaffilmark{9,14},
Evan Sinukoff\altaffilmark{2}
}
\shortauthors{Petigura et al.}

\altaffiltext{1}{California Institute of Technology, Pasadena, California, U.S.A. \href{mailto:petigura@caltech.edu}{petigura@caltech.edu}}
\altaffiltext{2}{Institute for Astronomy, University of Hawaii, 2680 Woodlawn Drive, Honolulu, HI, USA}
\altaffiltext{3}{Institute for Astronomy, University of Edinburgh, Blackford Hill, Edinburgh, EH9 3HJ}
\altaffiltext{4}{Lunar \& Planetary Laboratory, University of Arizona, 1629 E. University Blvd., Tucson, AZ, USA}
\altaffiltext{5}{NASA Exoplanet Science Institute, California Institute of Technology, 770 S. Wilson Ave., Pasadena, CA, USA}
\altaffiltext{6}{Astronomy Department, University of California, Berkeley, CA, USA}
\altaffiltext{7}{Department of Earth and Planetary Science, University of California Berkeley, Berkeley, CA 94720-4767, USA}
\altaffiltext{8}{Department of Physics \& Astronomy and Institute of Geophysics \& Planetary Physics, University of California Los Angeles, Los Angeles, CA 90095}
\altaffiltext{9}{NASA Ames Research Center, Moffett Field, CA, USA}
\altaffiltext{10}{Hubble Fellow}
\altaffiltext{11}{California Institute of Technology, Joint Center for Planetary Astronomy Fellow}
\altaffiltext{12}{NSF Graduate Research Fellow}
\altaffiltext{13}{NASA Sagan Fellow}
\altaffiltext{14}{NASA Postdoctoral Program Fellow}

\begin{abstract}
We report the discovery and confirmation of \stellar{name}{b} and c, two sub-Saturn planets orbiting a bright ($V$~=~11.3), metal-rich (\fe~=~\stellar{fe}~dex) \stellar{spectype} dwarf in the K2 Campaign 2 field. The planets are \planet{b}{Rp}~\Re and \planet{c}{Rp}~\Re and have orbital periods of \planet{b}{P}~d and \planet{c}{P}~d, near to the 2:1 mean-motion resonance. We obtained \stellar{nobsiodine} radial velocities (RVs) with Keck/HIRES and detected the reflex motion due to \stellar{name}{b} and c. These planets have masses of \planet{b}{mass}~\Me and \planet{c}{mass}~\Me, respectively. With low densities of \planet{b}{density}~\gcc and \planet{c}{density}~\gcc, respectively, the planets require thick envelopes of H/He to explain their large sizes and low masses. Interior structure models predict that the planets have fairly massive cores of \Mcoreb~\Me and \Mcorec~\Me, respectively. They may have formed exterior to their present locations, accreted their H/He envelopes at large orbital distances, and migrated in as a resonant pair. The proximity to resonance, large transit depths, and host star brightness offer rich opportunities for TTV follow-up. Finally, the low surface gravities of the \stellar{name} planets make them favorable targets for transmission spectroscopy by {\em HST}, {\em Spitzer}, and {\em JWST}.
\end{abstract}

\bibliographystyle{apj}

\section{Introduction}
The prime \Kepler mission (2009--2013) transformed our understanding of the prevalence and properties of extrasolar planets. In particular, statistical analyses showed that planets the size of Neptune and smaller vastly outnumber larger planets within 1~AU of GK dwarf stars \citep{Howard12,Fressin13,Petigura13b}. For example, 51\% of GK stars host a \Rp~=~1--4~\Re planet with $P$~=~5--100~d, while only 4.5\% of such stars host a \Rp~=~4--16~\Re planet in the same period range \citep{Petigura13b}.

\Kepler detected thousands of Earth-size and Sub-Neptune-size planets and a much smaller number of Jovians (\Rp~=~8--16~\Re) and sub-Saturns (\Rp~=~4--8~\Re) due to their comparative scarcity. Of these, only a small subsample orbit bright stars where follow-up observations such as radial velocity (RV) mass measurements and transmission spectroscopy are feasible. A major next step in exoplanet science is identifying transiting planets of all sizes orbiting bright stars.

Following the failure of two of the four reaction wheels onboard the {\em Kepler Space Telescope}, NASA began operating the telescope in a new mode called \ktwo \citep{howell:2014}. During \ktwo operations, the spacecraft observes a different region of the ecliptic plane every $\sim85$~d. By June 2016, \Kepler will have observed 10 additional fields in the \ktwo mode,  casting a wider net for planets around bright stars that are sparsely distributed on the sky.

\ktwo observations will improve our understanding of sub-Saturns. Around GK stars, sub-Saturns are almost twice as common as Jovians: 2.9\% of such stars host a sub-Saturn compared to the 1.6\% that host a Jovian \citep{Petigura13b}. Despite their relative abundance, few sub-Saturns have reliably measured masses and radii. The Exoplanet Orbit Database \citep{Han14}%
\footnote{
exoplanets.org, accessed 2015-08-24}
lists 13 sub-Saturns with density measured to 50\% or better compared to 174 Jovians. Ground-based transit surveys have a strong bias toward finding Jovian-size planets. In addition, Jovian-size planets have typical masses of $\sim$100--10,000~\Me vs. $\sim$10--100~\Me for sub-Saturns, making precise RV mass measurements more feasible for Jovians. 

Here we present the discovery of two sub-Saturn planets orbiting \stellar{name}. The planets have radii of \planet{b}{Rp}~\Re and \planet{c}{Rp}~\Re and  orbital periods of \planet{b}{P}~d and \planet{c}{P}~d, near the 2:1 mean-motion resonance. Their host star is a bright ($V$~=~11.3) \stellar{spectype} dwarf which allowed us to obtain precise RV mass constraints using Keck/HIRES. The planets have masses of \planet{b}{mass}~\Me and \planet{c}{mass}~\Me, respectively. We describe our photometric, imaging, and spectroscopic observations in Section~\ref{sec:observations}. In Section~\ref{sec:analysis}, we explain how we extract stellar and planet properties from our observations. In Section~\ref{sec:discussion}, we discuss the likely distribution in mass between core and envelope and how that relates to the planets' formation histories. We also investigate system dynamics in the context of long-term stability. We also place \stellar{name}{b} and c in the context of other sub-Saturns and discuss future follow-up opportunities. We give a brief summary in Section~\ref{sec:conclusions}.

\section{Observations}
\label{sec:observations}
\subsection{Discovery in \ktwo Photometry}
\stellar{name} was observed during \ktwo Campaign 2 with nearly continuous photometry from 2014 Aug 23 to 2014 Nov 13. The star is listed as EPIC-\stellar{epic} in the Mikulski Archive for Space Telescopes (MAST). It was selected for \ktwo observations based on \ktwo Guest Observer proposal GO2104 (PI:\ Petigura).  We list the star's identifying information, coordinates, and photometric properties in Table~\ref{tab:stellar}.  

\begin{figure}
\centering
\includegraphics[width=0.45\textwidth]{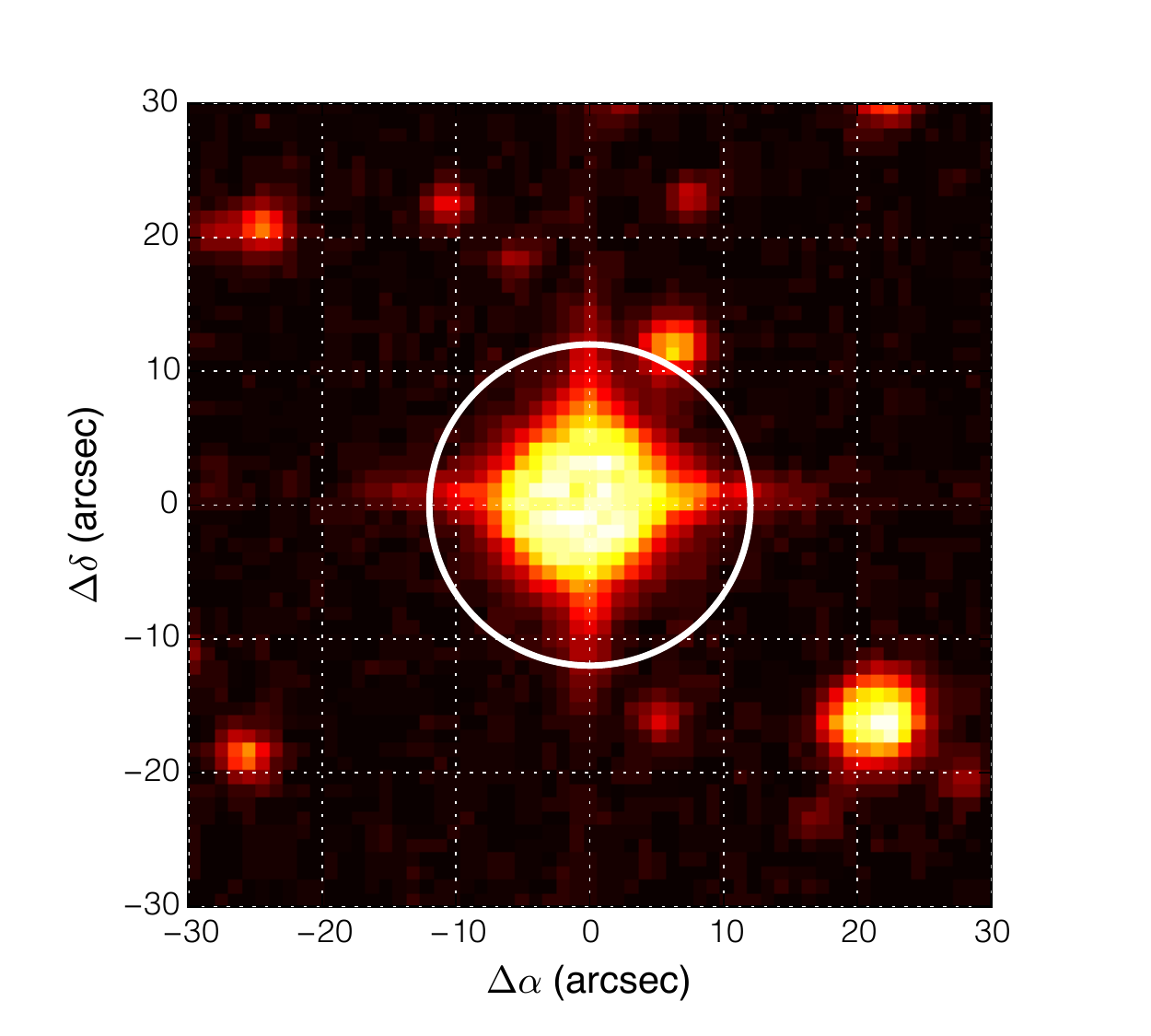}
\caption{POSS2 red planets observed in 1991. \protect\stellar{name} is in the center of the frame. The white circle shows the extent of the circular aperture used to extract the photometry of \protect\stellar{name}. No stars fall within our aperture that could dilute the light of \protect\stellar{name}. EPIC-203772026 sits just outside the boarder of our aperture. However, with $\Delta$\kepmag = 4.9 (EPIC catalog), it has negligible effect on the transit radius. EPIC-203772026 falls outside of the HIRES slit (width~=~0.86~arcsec). We rule out possibility that the observed transits are due to diluted eclipses of EPIC-203772026, because we observe the reflex velocities of \protect\stellar{name} due to planetary mass companions in our HIRES spectra (see Section~\ref{ssec:radial-velocities}).}
\label{fig:poss} 
\end{figure}

\begin{figure*}
\centering
\includegraphics[width=1\textwidth]{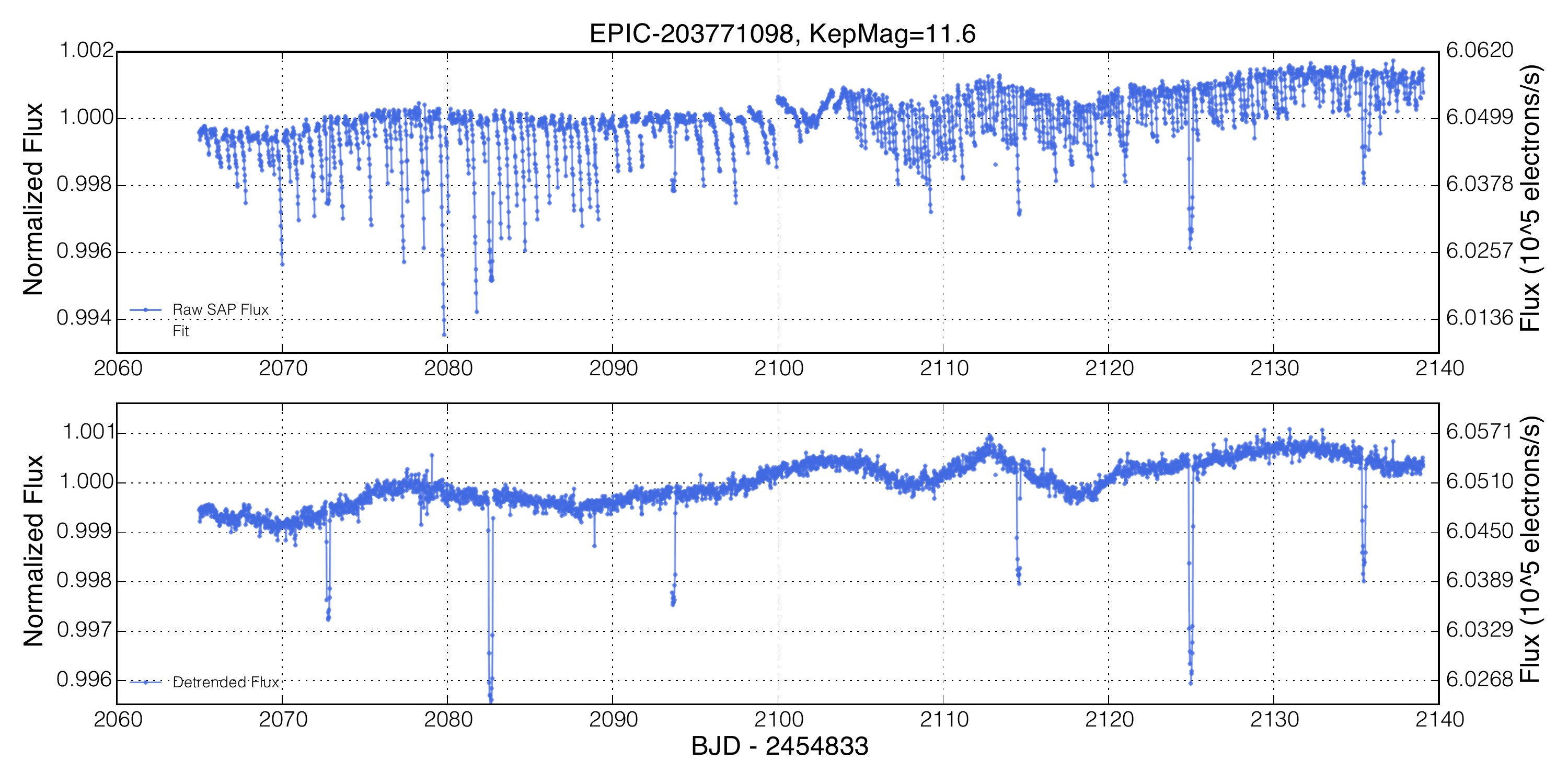}
\caption{{\em Top}: Raw photometry computed by summing the background-subtracted counts inside a circular aperture (3  pixel radius) centered on \protect\stellar{name}. {\em Bottom}: Photometry after correcting for variations due to telescope roll angle. Noise on three-hour timescales has been reduced by a factor of 8. The $\sim$0.1\% variability gives an upper limit to intrinsic stellar variability. Visual inspection gives a weak suggestion of a $\sim$20-day periodicity, but we do not consider this a compelling detection of rotational modulation. Since stars drift perpendicular to the roll direction over the course of a campaign, it is difficult to disentangle long-term astrophysical variability from position-dependent variability. The data used to produce the bottom panel is included as an electronic supplement.}
\label{fig:photometry} 
\end{figure*}

We extracted the photometry of \stellar{name} from the \Kepler pixel data, which we downloaded from the MAST. Our photometric extraction pipeline is described in \cite{crossfield:2015} and \cite{Petigura15}. In brief, during \ktwo operations the telescope is torqued by solar radiation pressure, causing it to roll around the boresight. This motion causes stars to drift across the CCD by $\sim$1~pixel every $\sim$6~hours. As stars sample different pixel-phases, inter-pixel sensitivity variations cause the apparent brightness of the star to change. We solve for the roll angle between each frame and an arbitrary reference frame. We model the time- and roll-dependent brightness variations using a Gaussian process. We also adjust the size of our circular extraction aperture to minimize the residual noise in the corrected light curve. This balances two competing effects: larger apertures yield smaller systematic errors while smaller apertures incur less background noise.  The circular extraction aperture ($r$~=~3~pixel) is shown in Figure~\ref{fig:poss}. Figure~\ref{fig:photometry} shows both the raw and corrected photometry for \stellar{name}. Our calibrated photometry is available as an online supplement.

We searched through the calibrated photometry using the \TERRA algorithm \citep{Petigura13b}. While two sets of transits are clearly visible by eye in the detrended \ktwo photometry for this star, we rely on \TERRA to search through the photometry of 10,000--20,000 light curves per \ktwo Campaign. After identifying the transits of planets b and c, we re-ran \TERRA on the photometry of \stellar{name} with the in-transit points removed and did not identify any additional transit candidates.

{\renewcommand{\arraystretch}{1.2}
\hspace{-1in}
\begin{deluxetable*}{l l l l }[bt]
\tabletypesize{\scriptsize}
\tablecaption{Stellar Parameters of \protect\stellar{name} \label{tab:stellar} }
\tablewidth{0pt}
\tablehead{
\colhead{Parameter} & Units & \colhead{Value} & \colhead{Source}
}
\startdata
\multicolumn{3}{l}{\hspace{1cm}Identifying Information} \\
EPIC ID       & --      & 203771098                     & EPIC \\
2MASS ID      & --      & 16101770-2459251                & 2MASS  \\
$\alpha$ R.A. & h:m:s   & 16:10:17.69                  & EPIC \\
$\delta$ Dec. & d:m:s   & $-$24:59:25.19               & EPIC \\
\multicolumn{3}{l}{\hspace{1cm}Photometric Properties} \\
\kepmag       & mag     & 11.65     & EPIC \\  
$B$             & mag     & $ 12.22 \pm 0.20  $ & APASS  \\
$V$             & mag     & $ 11.28 \pm 0.10  $ & APASS \\
$g^\prime$     & mag     & $ 13.99 \pm 0.90  $ & APASS \\
$r^\prime$     & mag     & $ 10.71 \pm 0.22  $ & APASS \\
$i^\prime$     & mag     & $ 10.64 \pm 0.01  $ & APASS \\
$J$             & mag     & $ 9.63 \pm 0.02  $ & 2MASS\\
$H$             & mag     & $ 9.29 \pm 0.02  $ & 2MASS\\
$Ks$            & mag     & $ 9.18 \pm 0.02  $ & 2MASS\\
\multicolumn{3}{l}{\hspace{1cm}Spectroscopic Properties} \\
\teff         & K             & $ 5743 \pm 60     $ & \SM, this paper \\ 
\logg         & dex           & $ 4.29 \pm 0.07     $ & \SM, this paper \\
\fe           & dex           & $ 0.42 \pm 0.04         $ & \SM, this paper \\ 
\vsini        & km~s$^{-1}$   &  $<2$                                           & \SM, this paper \\
$S_{\mbox{\scriptsize HK}}$  & --   &  0.128 & this paper \\
\logrhk       & dex   &  $-$5.26 & this paper \\
\multicolumn{3}{l}{\hspace{1cm}Derived Properties} \\
$\mu_{\alpha}$ & mas~yr$^{-1}$ &  $ -60.6 \pm 2.5   $ & \cite{zacharias:2012} \\
$\mu_{\delta}$ & mas~yr$^{-1}$  & $ -65.4 \pm 2.4 $ & \cite{zacharias:2012} \\
\Mstar        & \Msun   & $ 1.12 \pm 0.05     $ & \SM, \iso, this paper \\
\Rstar        & \Rsun   & $ 1.21 \pm 0.11     $ & \SM, \iso, this paper \\
\rhostar      & \gcc    & $ 0.89 \pm 0.23 $ & \SM, \iso, this paper \\
\Lstar        & \Lsun   & $ 1.44 \pm 0.33     $ & \SM, \iso, this paper \\
Distance      & pc      & $ 181 \pm 17           $ & \SM, \iso, this paper\\
Age           & Gyr     & 3.2--6.9        & \SM, \iso, this paper
\enddata
\tablecomments{\SM: \SpecMatch spectrum synthesis code \citep{Petigura15thesis}. \iso: \isochrones interface to the Dartmouth suite of stellar isochrones \citep{Morton15,Dotter08}.}
\end{deluxetable*}
}

\begin{figure*}
\begin{center}
\includegraphics[width=7in]{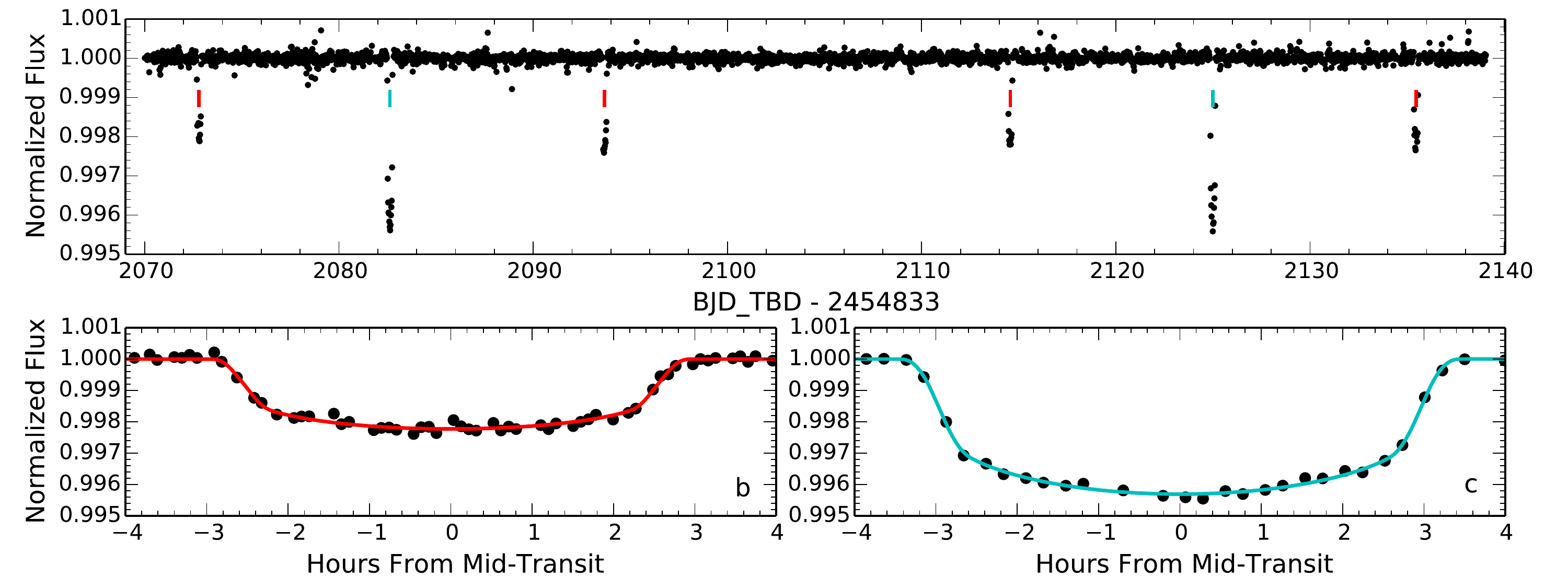}
\caption{{\em Top}: Calibrated \ktwo photometry for \protect\stellar{name}.  Vertical ticks indicate the times of transit. {\em Bottom}: Phase-folded photometry and best fit light curves for each planet. Best fit parameters from light curve fitting are tabulated in Table~\ref{tab:planet}.}
\label{fig:photometry-fits}
\end{center}
\end{figure*}

\subsection{Imaging}
\subsection{Archival and Adaptive Optics Imaging}
\label{sec:imageing}
We obtained near-infrared adaptive optics images of \stellar{name} using NIRC2 on the 10~m Keck II Telescope on the night of 2015-04-01 UT. We used the $1024\times1024$ NIRC2 array and the natural guide star system; the target star was bright enough to be used as the guide star. The data were acquired using the narrow-band Br-$\gamma$ filter using the narrow camera field of view with a pixel scale of 9.942 mas/pixel. The Br-$\gamma$ filter has a narrower passband (2.13--2.18 $\micron$), but a similar central wavelength (2.15 $\micron$) compared the Ks filter (1.95--2.34 $\micron$; 2.15 $\micron$) and allows for longer integration times before saturation. A 3-point dither pattern was utilized to avoid the noisier lower left quadrant of the NIRC2 array. The 3-point dither pattern was observed three times with 1 co-add and a 5.5 second integration time for a total on-source exposure time of $3\times3\times5.5$~s = 49.5~s.

\begin{figure}
\centering
\includegraphics[width=0.45\textwidth]{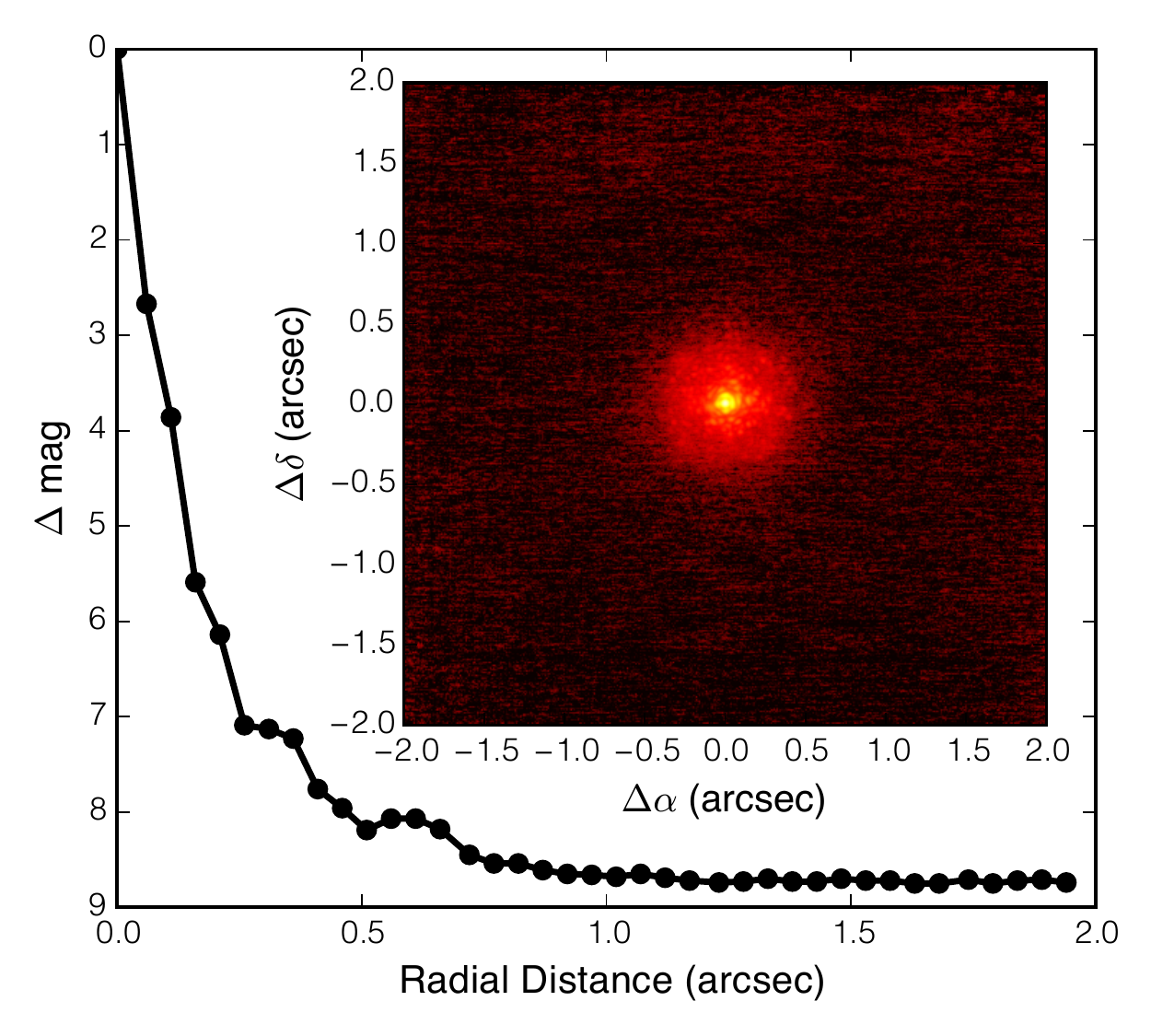}
\caption{NIRC2 K-band image and contrast curve. Our NIRC2 observations using the Br-$\gamma$ filter rule out companions having $K$-band contrasts of < 7.9 mag for separations of 0.7--8.0 arcsec. The inset shows a 4$\times$4 arcsec subregion in order to highlight the sensitivity of NIRC2 to companions at small orbital separations.}
\label{fig:ao} 
\end{figure}

The target star was measured with a resolution of 0.055~arcsec (FWHM). No other stars were detected within the 10~arcsec field of view of the camera. In the Br-$\gamma$ filter, the data are sensitive to stars that have $K$-band contrast of $\Delta K$ = 4.2 at a separation of 0.1 arcsec and $\Delta K$ = 7.9 at 0.5 arcsec from the central star. We estimate the sensitivities by injecting fake sources with a signal-to-noise ratio of 5 into the final combined images at distances of N~$\times$~FWHM from the central source, where N is an integer. Our combined NIRC2 image and contrast curve are shown in Figure~\ref{fig:ao}.

\subsection{Spectroscopy}
We observed \stellar{name} with the High Resolution Echelle Spectrometer (HIRES; \citealt{Vogt94}) on the 10~m Keck Telescope I. Between 24 June and 3 October 2015, we obtained \stellar{nobsiodine} spectra through an iodine cell mounted directly in front of the spectrometer slit. The iodine cell imprints a dense forest of absorption lines which serve as a wavelength reference. We also obtained a ``template'' spectrum without iodine.  We used an exposure meter to achieve a constant signal to noise ratio of 110 per HIRES pixel on blaze near 550~nm.  Exposure times were in the range 6--12 min.  RVs were determined using standard procedures of the California Planet Search \citep[CPS;][]{howard:2010b} including forward modeling of the stellar and iodine spectra convolved with the instrumental response \citep{Marcy92,Valenti95}. The radial velocities are tabulated in Table~\ref{tab:rv}. We also list the measurement uncertainty of each RV point, which range from 1.5 to 2.0~\ms from the uncertainty on the mean RV of the $\sim$700 spectral chunks used in the RV pipeline. 

We measured the strength of the \ion{Ca}{2} H \& K lines and found that \protect\stellar{name} is an inactive star.  We see no emission reversal in the cores of these lines.  Table~\ref{tab:stellar} lists the median values of $S_{\mbox{\scriptsize HK}}$ using the method of \cite{Isaacson10} and $\log R'_{\mbox{\scriptsize HK}}$ computed using $B-V = 0.673$ estimated from \teff according to the relation from \cite{Valenti05}.

We searched for companions with separations smaller than $\approx0.1$~arcsec, where our sensitivity to sources from AO imaging declines (see Figure~\ref{fig:ao}). Adopting the methodology in \cite{kolbl:2015}, we searched for spectroscopic binaries in our HIRES spectrum. We detect no secondary set of lines from a star having $\Delta V < 5$~mag shifted by more than 15 km/s relative to the lines of the primary star. Shifts of $\Delta v \gtrsim$~15~km/s correspond to orbital separations of $\lesssim4$~AU. 

\section{Analysis}
\label{sec:analysis}
\subsection{Stellar Properties}
\label{ssec:stellar-properties}
We analyzed our iodine-free template spectrum from HIRES using the \SpecMatch spectrum synthesis code \citep{Petigura15thesis}. \SpecMatch is a general tool for extracting stellar \teff, \logg, \fe, and \vsini by fitting high-resolution spectra. \SpecMatch generates synthetic spectra at arbitrary \teff, \logg, \fe, and \vsini by interpolating between LTE models of \cite{Coelho05} and applying broadening kernels that account for line broadening due to stellar rotation and macroturbulence and the instrumental profile of the spectrometer. \teff, \logg, \fe, and \vsini are adjusted in order to yield the best-matching spectrum in $\chi^{2}$ sense. We determined that \stellar{name} is a metal-rich G3 star having \teff~=~\stellar{teff}~K, \logg~=~\stellar{logg}~dex, \fe~=~\stellar{fe}~dex, and \vsini < 2~\kms. Our uncertainties in \teff and \fe are based on comparisons with touchstone stars in the literature with stellar parameters from asteroseismology \citep{Huber13}, LTE-modelling \citep{Valenti05,Torres12}, and Rossiter-McLaughlin measurements \citep{Albrecht12}.

We converted spectroscopic parameters into physical stellar properties using the \isochrones python package \citep{Morton15}, which provides a convenient interface to the Dartmouth suite of stellar isochrones \citep{Dotter08}. \stellar{name} is slightly larger and more massive than the Sun: \Mstar~=~\stellar{Mstar}~\Msun and \Rstar~=~\stellar{Rstar}~\Rsun. We list the spectroscopic and derived physical properties of \stellar{name} in Table~\ref{tab:stellar}.

{\renewcommand{\arraystretch}{1.4}
\begin{deluxetable*}{l l l l l}
\tabletypesize{\scriptsize}
\tablecaption{  Planet Parameters \label{tab:planet}}
\tablewidth{0pt}
\tablehead{
\colhead{Parameter} & \colhead{Units} & \colhead{b} & \colhead{c} 
}
\startdata
\multicolumn{3}{l}{\hspace{1cm} Light curve fitting } \\
   $T_{0}$ & BJD$_\mathrm{TDB} - 2454833$ & $2072.7948\pm0.0007$ &$2082.6251\pm0.0004$ \\
       $P$ &          d & $20.8851\pm0.0003$ &$42.3633\pm0.0006$ \\
       $i$ &        deg & $89.25_{ -0.61 }^{ +0.49 }$ &$89.76_{ -0.21 }^{ +0.18 }$ \\
 $R_P/R_*$ &         \% & $4.31_{ -0.08 }^{ +0.17 }$ &$5.94_{ -0.04 }^{ +0.10 }$ \\
  $T_{14}$ &         hr & $5.48_{ -0.04 }^{ +0.07 }$ &$6.47_{ -0.03 }^{ +0.04 }$ \\
  $T_{23}$ &         hr & $4.95_{ -0.11 }^{ +0.05 }$ &$5.70_{ -0.06 }^{ +0.03 }$ \\
   $R_*/a$ &         -- & $0.035_{ -0.002 }^{ +0.005 }$ &$0.019_{ -0.000 }^{ +0.001 }$ \\
       $b$ &         -- & $0.37_{ -0.24 }^{ +0.22 }$ &$0.22_{ -0.16 }^{ +0.17 }$ \\
\rhostarcirc &     \gcc & $1.00_{ -0.33 }^{ +0.21 }$ &$1.47_{ -0.23 }^{ +0.31 }$ \\

       $a$ &         AU & $0.154\pm0.002$ &$0.247\pm0.004$  \\
     \Sinc &        \Se & $60\pm14$ &$24\pm5$ \\
      \teq &          K & $767\pm177$ &$606\pm139$ \\
       \Rp &        \Re & $5.68\pm0.56$ &$7.82\pm0.72$ \\

\\
\multicolumn{3}{l}{\hspace{1cm} Circular RV model (adopted)} \\
       $K$    & \ms     & $4.5\pm1.1$ &$4.6\pm1.2$ \\
     $\gamma$ & \ms     & \multicolumn{2}{c}{ $-2.5 \pm 0.9$  } \\
     \dvdt    & \msyr   & \multicolumn{2}{c}{ $-23.9 \pm 9.7$  } \\
    \sigjit   & \ms     & \multicolumn{2}{c}{ $3.4 \pm 0.7$  } \\
       \Mp    & \Me     & $21.0\pm5.4$ &$27.0\pm6.9$ \\ 
    $\rho$    & g cm$^{-3}$ & $0.63\pm0.25$ &$0.31\pm0.12$ \\
\\
\multicolumn{3}{l}{\hspace{1cm} Eccentric RV model} \\
       $K$    & \ms     & $5.1\pm1.2$ &$5.3\pm1.1$ \\
     \ecosw   & --      & $0.20\pm0.09$ &$0.00\pm0.09$ \\   
     \esinw   & --      & $-0.06\pm0.16$ &$-0.02\pm0.15$ \\   
     $e$      & --      & $0.24_{-0.11}^{+0.11}$ & < 0.39 (95\%) \\   
     $\gamma$ & \ms     & \multicolumn{2}{c}{ $-2.7 \pm 1.0$  } \\
     \dvdt    & \msyr   & \multicolumn{2}{c}{ $-22.5 \pm 9.2$  } \\
    \sigjit   & \ms     & \multicolumn{2}{c}{ $2.9 \pm 0.6$  } \\
       \Mp    & \Me     & $23.2\pm5.3$ &$31.0\pm6.4$ \\ 
    $\rho$    & g cm$^{-3}$ & $0.70\pm0.26$ &$0.36\pm0.12$ \\
\enddata
\end{deluxetable*}
}

\subsection{Light curve modeling}
\label{ssec:light-curve-modeling}
We analyzed \ktwo transit light curve using the same approach described by \cite{crossfield:2015}. In brief, we fit each planet's transit
separately using a minimization and Markov Chain Monte Carlo (MCMC) analysis \citep{foreman-mackey:2012} using the {\tt batman} code \citep{Kreidberg15} to model the light curves that assumes a linear transit ephemeris for each planet.

When modeling the transit photometry, we adopt a quadratic limb-darkening law. We used the {\tt LDTk} Limb Darkening Toolkit \citep{Parviainen15} to derive limb-darkening coefficients of $u_1$~=~0.568$\pm$0.003 $u_2$~=~0.098$\pm$0.005. We doubled the uncertainties associated with the limb-darkening parameters and incorporated them as Gaussian priors in the MCMC light curve analysis. All of the MCMC parameters show unimodal distributions. The transit profiles alone do little to constrain orbital eccentricity and give upper limits on $e_b < 0.78$ and $e_c < 0.81$ at 95\% confidence. Figure~\ref{fig:photometry-fits} shows the \stellar{name} photometry and best fit models, and Table~\ref{tab:planet} summarizes the final values and uncertainties.

The transit profile constrains the mean stellar density if one assumes a circular orbit. Since we fit each planet separately, we obtain two independent measurements for \rhostarcirc, \planet{b}{rhostar}~\gcc and \planet{c}{rhostar}~\gcc. In addition we also have a spectroscopic estimate of \rhostar~=~\stellar{rho}~\gcc. All three estimates of mean stellar density are consistent at the 2-$\sigma$ level. In our analysis, we have modeled the light curve as a single unblended star. While our AO and spectroscopic observations rule out stars with $\Delta Kp \lesssim 5$ inside $\sim$4~AU and outside $\sim$20~AU, we have not covered parameter space entirely. There is a small possibility that our transit profiles could be diluted by an additional star, affecting primarily the derived planet radii. However, we confirm these planets without the need for statistical validation with RVs as described in the following section.

\subsection{Radial Velocities}
\label{ssec:radial-velocities}
We detected RV variability matching the orbital periods and phases of \stellar{name}{b} and c that were measured from the \ktwo light curve.  Measuring the masses of these planets (as described below) confirms their existence and rules out false positive scenarios.  

We modeled the stellar RV time series as the sum of two Keplerian orbits. We considered both eccentric and circular orbits. Circular orbits require three parameters per planet: orbital period $P$, time of transit $T_{0}$, and the Doppler semi-amplitude $K$. In addition, we allowed for an arbitrary RV offset, $\gamma$, and a linear acceleration, \dvdt. To assess the quality of a given model we evaluated the log-likelihood, \loglike, according to the prescription given in \cite{Howard14}. This likelihood definition incorporates RV ``jitter'' (\sigjit), an additional RV uncertainty due to astrophysical and instrumental sources. To guard against non-physical values of $K$ and \sigjit, we parametrized the model using $\log K$ and $\log \sigjit$. We imposed no prior on $\log \sigjit$. Because $P$ and $T_0$ are measured with exquisite precision from the \ktwo photometry, we held these parameters fixed during our RV analysis. While the \stellar{name}{bc} pair's proximity to resonance will result in strong dynamical interactions, we expect TTVs on the order of $\sim${3--6}~hr, not an appreciable fraction of an orbital period (see Section~\ref{sssec:ttv}). Therefore, we do not consider departures from strict Keplerian motion when modeling the RVs.

\begin{figure*}
\centering
\includegraphics[width=0.8\textwidth]{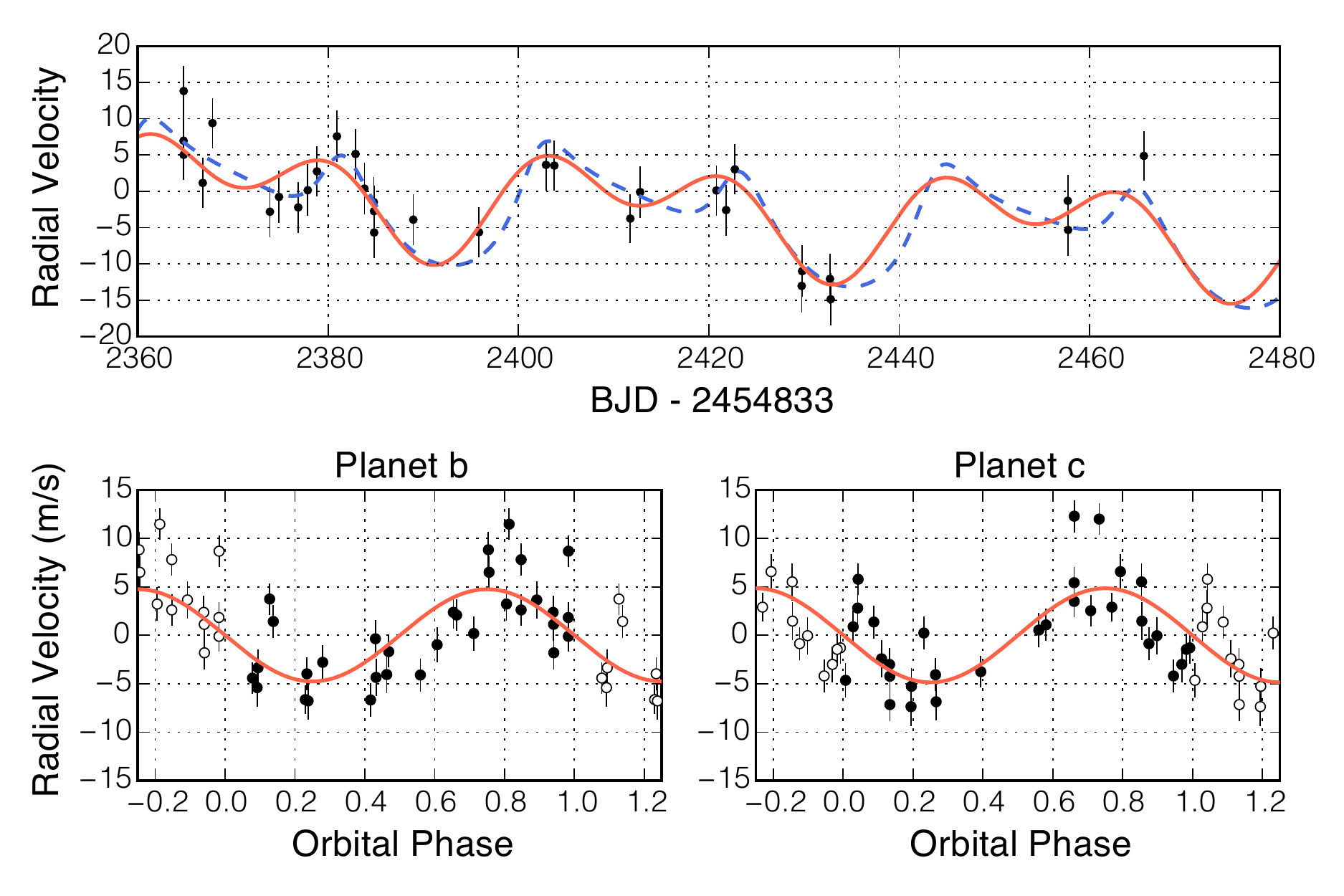}
\caption{The top panel shows the radial velocity time series collected using Keck/HIRES between 24 June 2015 and 03 October 2015. The red line shows the best fit Keplerian, assuming a circular orbit. The blue dashed line shows the best fit Keplerian allowing for eccentricity to vary. While the eccentric solution has higher likelihood than the circular solution $\ln \mathcal{L}$~=~$-85.3$ vs $-80.1$, it comes at the expense of more free parameters. We adopt the circular solution as our system parameters. The bottom panels show the RVs folded on the ephemerides of planet b and c. In these plots, the contribution of the other planet as well as the contribution from the trend (parametrized by $\gamma$ and \dvdt) has been removed.}
\label{fig:rv} 
\end{figure*}

The red curve in Figure~\ref{fig:rv} shows the maximum likelihood model which we found using the Limited-Memory Broyden-Fletcher-Goldfarb-Shanno optimization routine \citep{Byrd95} as implemented in the {\tt scipy} Python package \citep{Jones01}. The bottom panels show the maximum likelihood models for each planet individually. We explored the likelihood surface using MCMC as implemented in the {\tt emcee} Python package \citep{Foreman-Mackey13}. Table~\ref{tab:planet} summarizes the median posterior values and the 14\% and 86\% quantiles. We detected the reflex velocities due to both planets. Assuming circular orbits, \stellar{name}{b} and \stellar{name}{c} have masses of \planet{b}{mass}~\Me and \planet{c}{mass}~\Me, respectively. 

Eccentric models included two additional parameters per planet: $e$, eccentricity and \lonperi, the longitude of periastron of the star's orbit. Following \cite{Eastman13}, we re-parametrized $e$ and $\lonperi$ as \sqrtecosw and \sqrtesinw, which mitigates the Lucy-Sweeney bias toward non-zero eccentricity \citep{Lucy71}. The maximum likelihood model is shown as a blue dashed curve in Figure~\ref{fig:rv}, and the MCMC posteriors are summarized in Table~\ref{tab:planet}. When we included eccentricity in the models, the planets have masses \planetecc{b}{mass}~\Me and \planetecc{c}{mass}~\Me, respectively. Interestingly, our eccentric models predict $e_b$~=~\planetecc{b}{e} while $e_c$ is consistent with zero (< 0.39 at 95\% confidence).

We assessed the relative merits of the eccentric and circular models using the Bayesian Information Criterion (BIC). The BIC is defined as $\mathrm{BIC} = -2 \ln \mathcal{L_{\mathrm{max}}} + k \ln N$ where $\mathcal{L_{\mathrm{max}}}$ is the maximized likelihood, $k$ is the number of free parameters, and $N$ is the number of observations \citep{Schwartz78,Liddle04}. For our RV time series $N$~=~\stellar{nobsiodine}. When comparing two models, the model with the lower BIC is preferred. The BIC penalizes models with low likelihood and high complexity. BIC(circular)~$-$~BIC(eccentric)~=~$-4.9$. Because the best fit circular model has lower BIC, we adopt its associated best fit parameters as our preferred system parameters. However, we discuss the dynamical implications of eccentric orbits in Section~\ref{ssec:dynamics}.

We also observe a linear trend in the radial velocities of $-23.9 \pm 9.7$~\msyr. This trend is marginally significant and could indicate an additional body in the system. Following \cite{Winn10}, we consider the range of possible $\Mp \sin i$ and $a$ that could produce the observed trend. A body on a circular orbit with $\Mp \ll \Mstar$ induces a reflex acceleration of $\dvdt = G \Mp / a^{2}$. The mass and separation of the planet (or star) is given by 
\[
\Mp \sin i \sim 42\, \Me \left( \frac{a}{1 \mathrm{AU} } \right)^{2}. 
\]

We advocate for continued RV monitoring of \stellar{name} to determine whether the observed RV trend is due to an additional long-period planet.

Our circular orbital solution favors an RV ``jitter'' (\sigjit) of 3.4~\ms. Here, we assess whether that jitter is consistent with the ensemble of stars monitored with precision RVs. \cite{Isaacson10} derived an empirical relationship between \logrhk, B-V color, and \sigjit. The relationship in \cite{Isaacson10} predicts that at star with $B-V$~=~0.65%
\footnote{
Because the light of \stellar{name} suffers significant extinction, we derive $B-V$~=~0.65 from our \SpecMatch/\isochrones analysis, as opposed to using the $B$ and $V$ magnitudes from the EPIC catalog.
}
and \logrhk~=~$-$5.26 will have $\sigjit \approx 2.0$~\ms. One explanation for the higher-than-expected jitter is the presence of additional short-period planets having $K$~$\approx$~1--2~\ms. Again, we encourage additional RV monitoring of \stellar{name} to search for additional short-period planets not detected in the K2 photometry. The detection such a planet would not only add to the dynamical richness of the \stellar{name} system, but would also lead to better constraints on the orbital parameters of \stellar{name}{b} and c.

\begin{deluxetable}{l r r}[bt]
\tabletypesize{\scriptsize}
\tablecaption{ Relative Radial Velocities \label{tab:rv}} 
\tablewidth{0pt}
\tablehead{
\colhead{BJD - 2454833} & \colhead{Radial Velocity} & \colhead{Uncertainty} \\
                        & \colhead{\ms}             & \colhead{\ms}
}
\startdata
2364.81958 & $6.96$ & 1.59\\2364.82510 & $5.02$ & 1.60\\2364.83070 & $13.81$ & 1.66\\2366.82758 & $1.15$ & 1.65\\2367.85265 & $9.39$ & 1.64\\2373.88815 & $-2.82$ & 1.72\\2374.85241 & $-0.77$ & 1.91\\2376.86382 & $-2.22$ & 1.71\\2377.86607 & $0.15$ & 1.84\\2378.83401 & $2.74$ & 1.65\\2380.93080 & $7.57$ & 1.86\\2382.88614 & $5.14$ & 1.68\\2383.82353 & $0.37$ & 1.90\\2384.79994 & $-1.48$ & 1.69\\2384.82899 & $-2.74$ & 1.68\\2384.83972 & $-5.68$ & 1.72\\2388.95596 & $-3.91$ & 1.71\\2395.85726 & $-5.64$ & 1.64\\2402.89876 & $3.64$ & 1.76\\2403.77132 & $3.54$ & 1.65\\2411.75570 & $-3.75$ & 1.46\\2412.79420 & $-0.11$ & 1.78\\2420.80302 & $0.11$ & 1.64\\2421.82280 & $-2.59$ & 1.76\\2422.74212 & $3.02$ & 1.66\\2429.76175 & $-13.03$ & 1.98\\2429.81023 & $-11.00$ & 1.88\\2432.73232 & $-12.06$ & 1.70\\2432.80724 & $-14.87$ & 1.91\\2457.71690 & $-1.31$ & 1.93\\2457.75480 & $-5.32$ & 1.94\\2465.71074 & $4.87$ & 1.62
\enddata
\end{deluxetable}

\section{Discussion}
\label{sec:discussion}
\subsection{Core-Envelope Structure}
\stellar{name}{b} \& c are among only a handful of transiting sub-Saturns with well-measured masses. With two sub-Saturns in the same system, we have a rare chance to compare the possible compositions of these planets to each other and to the general population of sub-Saturns. We examine possible compositions with the interior and thermal evolution models of \cite{Lopez14} which track the cooling and contraction of planets with H/He envelopes and allow us to convert measured masses, radii, and incident fluxes of \stellar{name}{b} \& c into estimates of H/He mass fraction. 

We modeled planets with solar metallicity H/He envelopes atop a fully differentiated Earth composition core.  According to these models, we find that \stellar{name}{b} is \fenvb{\%}  H/He by mass, while \stellar{name}{c} is \fenvc{\%} H/He by mass. \stellar{name}{b} \& c then have core masses of \Mcoreb~\Me and \Mcorec~\Me, respectively. The uncertainty on the envelope fraction includes the observational uncertainties on planet mass, radius, age, and incident flux along with theoretical uncertainties such as the iron fraction and heat capacity of the rocky core. Our uncertainty in envelope mass is dominated by planet radius errors; our uncertainty in planet core mass is dominated by uncertainties in planet mass.

These planets are sufficiently large that our conclusions are insensitive to variations in the core composition. While pure water cores are likely unphysical, we repeated the above calculations for planets with 98\% water cores in order to set a lower bound on the H/He envelope fraction. Using these models of \stellar{name}{b} \& c, we found envelope mass fractions of \fenvwaterb \& \fenvwaterc and core masses of \Mcorewaterb~\Me and \Mcorewaterc~\Me, respectively. The effect of changing the assumed core composition from Earth-like to pure water has a small effect on the derived core masses (within the statistical uncertainties). We adopt \Mcoreb~\Me and \Mcorec~\Me as the core masses of \stellar{name}{b} \& c, respectively.

\subsection{Formation Scenarios}
\label{ssec:formation-scenarios}
The inferred core-envelope structures of \stellar{name}{b} \& c pose some challenges to explaining their formation. How did \stellar{name}{c} end up with twice as much gas as \stellar{name}{b} despite forming in the same disk with a similar core mass? Another challenge is explaining how \stellar{name}{c} is composed of half H/He gas, but somehow avoided runaway accretion as predicted in standard models of core accretion (e.g., \citealt{Pollack96,Lee14}).

While the different densities of planet pairs like Kepler-36b \& c can be understood in terms of differing XUV-driven mass loss histories \citep{Lopez13}, mass loss likely played only a minor role for \stellar{name}{b} \& c. The planets are only modestly irradiated, and their cores are relatively massive compared to typical hot sub-Neptunes. Using the coupled thermal evolution and photo-evaporation model of \cite{Lopez13}, we find that both planets would have only been $\approx$1\% more massive at an age of 10 Myr.

The fact that the \stellar{name}{bc} pair are near the 2:1 mean-motion resonance suggests they formed at larger orbital separations and experienced convergent inward migration (e.g., \citealt{Murray00}; see also \citealt{Deck15}). Formation at $\gtrsim1$~AU as opposed to their current locations ($\sim$0.2~AU), together with inward migration, could explain the large inferred envelope fractions \citep{Lee15b}. \cite{Lee15a} derived analytic scaling relations for atmospheric accretion. For planets at $\gtrsim$ 1~AU with dust-free atmospheres, they found that the gas-to-core mass ratio scales as
$M_{\mathrm{core}}^{1} T_{\mathrm{eq}}^{-1.5}$ 
(equation 24 of their paper), with
$T_{\mathrm{eq}}$ equal to the equilibrium
surface temperature. The dependence on $T_{\mathrm{eq}}$ arises because colder planets have lower opacities and therefore cool and accrete faster. Since \stellar{name}{c} presumably formed exterior to \stellar{name}{b} and had a lower $T_{\mathrm{eq}}$, these dust-free accretion models may explain, in a natural way, why the outer planet in the \stellar{name}{bc} pair has a more massive envelope.

In the case of dusty atmospheres, \cite{Lee15a} found that the accreted gas fraction is independent of local disk temperature. Thus if the early atmospheres of \stellar{name}{b} \& c were dusty, different formation locations could not alone explain their different envelope masses. Dusty envelopes might still be accommodated if \stellar{name}{b} initially formed with a more massive envelope but lost a large fraction of it to a giant impact. Recent studies of giant impacts have found that a collision with an equal mass impactor can reduce a planet's gas fraction by a factor of $\sim$2 \citep{Liu15,Inamdar15}, thus providing an alternative explanation for the difference between the inferred gas fractions of \stellar{name}{b} and c.

With this one system, it is not possible to distinguish between the formation hypotheses of dust-free gas accretion vs.\ giant impacts. However, it is intriguing that the longer period planet has the much larger envelope fraction, as we might expect from dust-free gas accretion and convergent migration \citep{Lee15b}. As more sub-Saturns are found and characterized, ensemble properties should shed light on their formation. If stochastic processes like giant impacts determine envelope masses, we expect no correlation between orbital distance and envelope fraction. However, if gas accretion is governed by local disk properties, we should see correlations with orbital distance---as is arguably already observed by the well-known increase in the occurrence rate of Jupiter-mass gas giants beyond $\sim$1 AU \citep{Cumming08}.

Explaining how \stellar{name}{c} roughly doubled in mass while accreting gas, yet somehow avoided runaway accretion is difficult in the context of standard models of core accretion (e.g., \citealt{Pollack96,Lee14}). The \citet{Lee15b} scenario of dust-free gas accretion in a disk coupled with inward migration raises a concern of fine tuning since it requires 
that \stellar{name}{c} reach the threshold of runaway (its gas-to-core ratio is modeled to be \fenvc{\%}) without actually running away.
However, we note that sub-Saturn-sized planets are not common outcomes of planet formation (2.9\% of GK stars host a sub-Saturn with $P$~<~100~d; \citealt{Petigura13b}) and near-resonant sub-Saturns are rarer still.

\subsection{System Dynamics}
\label{ssec:dynamics}
Given the current dataset, we are hesitant to claim a non-zero eccentricity for planet b. However, our eccentric model does have some precedent among previously discovered systems. GJ876c and b have orbital periods of 30.08~d and 61.12~d respectively, and, like \stellar{name}{b} and c, lie just outside 2:1 mean-motion resonance \citep{Marcy98,Delfosse98,Marcy01}. N-body fits to the GJ876 RVs show that planet c is moderately eccentric ($e = 0.25591\pm0.00093$) while planet b has a nearly circular orbit ($e = 0.0292\pm0.0015$; \citealt{Rivera10}).  The high precision eccentricity measurements in this case are from the large Doppler amplitudes ($\sim$100 \ms) and the detection of resonant interactions.

We consider here the dynamical implications of eccentric orbits, assuming the system has a long-lived orbital architecture. The system dynamics are governed by the two planets' masses, eccentricities, and longitudes of pericenter, $\varpi$.%
\footnote{We use $\varpi$ to refer to the planet's orbit as opposed to \lonperi, which refers to the star's orbit.}
Instead of performing a uniform exploration of this six-dimensional parameter space, we consider systems drawn from our MCMC exploration of eccentric RV solutions. First, roughly 25$\%$ of the models in our MCMC chain satisfy $a_c(1-e_c)<a_b(1+e_b)$. Given that the distribution of planet $\varpi$ is nearly uniform, many of these solutions correspond to crossing orbits.  

We also considered whether the systems are Hill stable, using the full criterion, which is based on conservation of the quantity $L^2E$, where $L$ and $E$ are the total orbital angular momentum and energy of the system \citep{Marchal,Milani}. Roughly half of the MCMC realizations fail the Hill criterion. These include all solutions with $e_b \gtrsim 0.3$ or $e_c \gtrsim 0.3$. While eccentricity of planet b ($e_b$~=~\planetecc{b}{e}) is likely less than 0.3, the 1-$\sigma$ confidence interval extends past 0.3. In this case, the system must be in some type of resonant phase protection which prevents close approaches in order to be long-lived \citep[e.g.,][]{Gladman,Barnes}. 
On the other hand, orbits which satisfy the Hill criterion, though protected from collisions, are not necessarily long-lived as weak encounters can still lead to large and erratic variations in the orbital elements. To test this, we selected 100 planet masses, orbital eccentricities and longitudes of pericenter randomly from the MCMC chain. We integrated these initial conditions using a Wisdom-Holman mapping with a symplectic corrector employed \citep{WisdomHolman,SymplecticCorrector}. Our timestep was 0.25 days, and we ensured that the fractional energy conservation was high (typically $\sim 10^{-10}$). The integrations lasted for $10^6$ years, or $\approx$20 million orbits of the inner planet. 

Although roughly half of the orbits failed the Hill criterion, only 9 showed instability during the integrations (deviations in semimajor axes larger than $5\%$ of the initial values).  To understand why, we selected the orbits which failed the Hill criterion yet remained long-lived, and looked at the orbital evolution of the eccentricities and the angle $\Delta \varpi = \varpi_b-\varpi_c$ on timescales of $3\times10^4$ orbits of the inner planet. Roughly 70$\%$ show apparently regular evolution, and the majority exhibit libration of $\Delta \varpi$, about zero or $\pi$, indicating a resonant protection mechanism. Although there is no preferred value of $\Delta \varpi$ based on the RV data, at high eccentricities the 2:1 resonance is wide, and so it is not surprising that many are in resonance, albeit with large libration amplitudes.

The majority of the remaining 30$\%$ of the orbits appear chaotic, with erratic variation of eccentricities and alternations between circulating and libration of $\Delta \varpi$ and/or the (mean-motion) resonant angles $2\lambda_c-\lambda_c-\varpi_b$ and $2\lambda_c-\lambda_c-\varpi_c$. It is interesting that despite this chaotic behavior, the effective lifetimes of these orbits are relatively long. We expect longer integrations would reveal unstable behavior. 

In conclusion, this limited look into the long-term stability of the orbital solutions to the RV data suggests that orbits with large eccentricities are plausible, even if the system fails the Hill criterion, if the system is in resonance. 

\subsection{Sub-Saturn Planets}
\begin{figure*}
\centering
\includegraphics{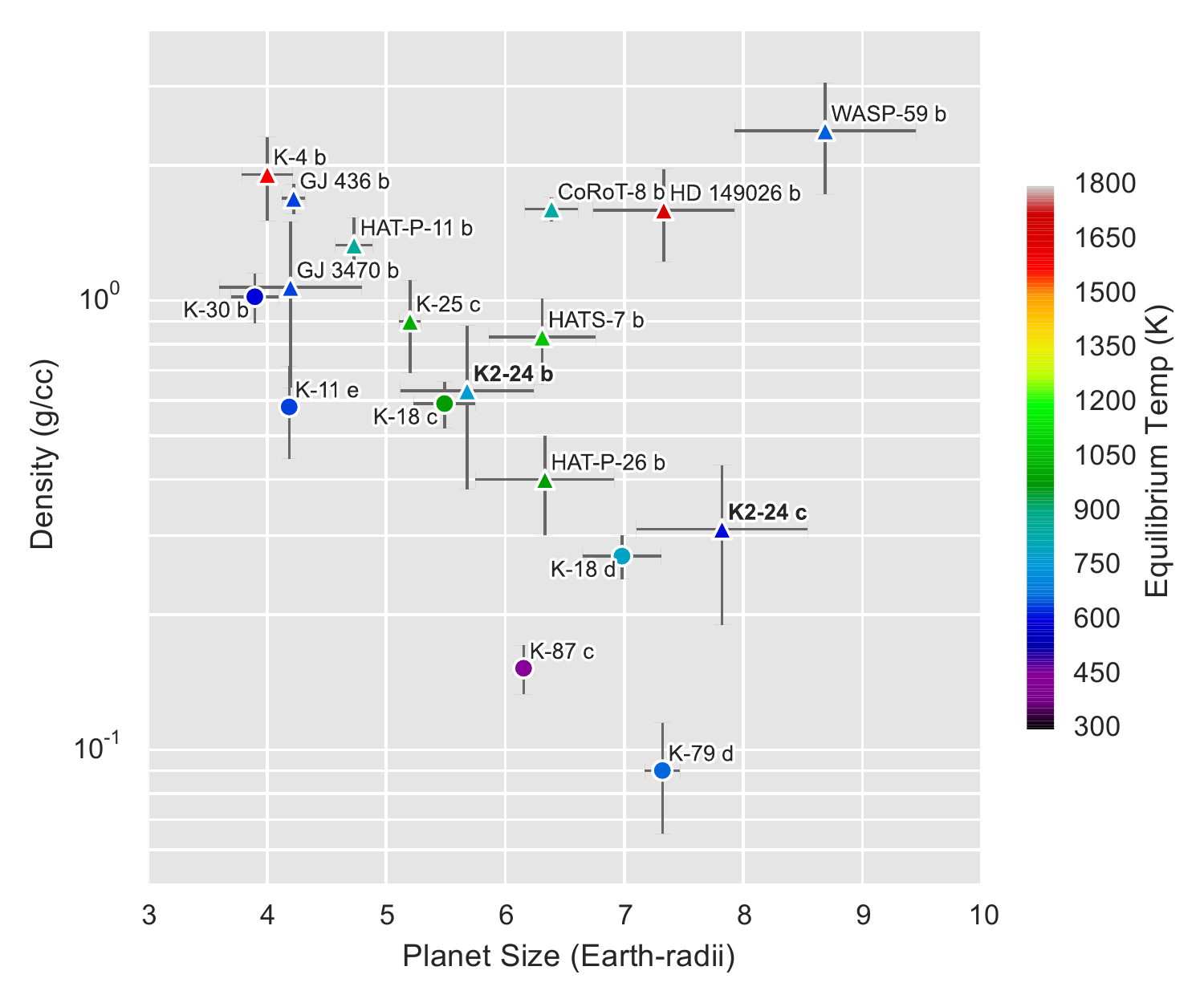}
\caption{Planet radii and densities for planets having \Rp~=~4--8~\Re where density is measured to better than 50\%. We have also included those planets with one-sigma radius measurements consistent with 4--8~\Re. The symbol color represents the (zero-albedo) equilibrium temperature. The symbol shape indicates the observational technique used to measure planet mass. Triangles and circles represent RV and TTV measurements, respectively. Planets taken from the Exoplanet Orbit Database and from \cite{Bakos15}. Note: planets from the \Kepler prime mission are designated with ``K'' (e.g. K-79d is Kepler-79d)}
\label{fig:Rp-rho} 
\end{figure*}

Here, we put the \stellar{name} system in the context of the other sub-Saturns. Figure~\ref{fig:Rp-rho} shows the densities and radii of planets having \Rp~=~4--8~\Re and density measured to better than 50\%, i.e.\ $\sigma(\rho) / \rho < 50\%$. The symbol colors represent the planet equilibrium temperature assuming zero albedo, and the symbol shapes indicate whether TTVs or RVs were used to measure planet mass. The \stellar{name} planets are labeled in bold. The \stellar{name} planets are fairly typical compared other sub-Saturn planets. 

The relative sizes and densities of the \stellar{name}{bc} pair are reminiscent of the Kepler-18cd pair. Kepler-18c has a mass of $17.3\pm1.9$~\Me, radius of $5.49\pm0.26$~\Re, and a density of 0.59~\gcc, similar to \stellar{name}{b}. Kepler-18d has a mass of $16.4\pm1.4$~\Me, a radius of $6.98\pm0.33$~\Re, and a density of $0.27\pm0.03$~\gcc. While Kepler-18b is smaller and less massive than \stellar{name}{c}, it has a similar density. Kepler-18cd also lie near the 2:1 mean-motion resonance \citep{Cochran11}. 

While there are still relatively few sub-Saturns with well-measured masses and radii, there are some trends worth noting. Densities measured from TTVs tend to be lower than RV-measured densities. This trend was noted by \cite{weiss:2014} for planets smaller than 4~\Re. Here, we offer some observational and astrophysical explanations. The Doppler semi-amplitude, $K$, depends primarily on planet mass, and has a weaker dependence on orbital period, eccentricity, and stellar mass. The TTV technique is also sensitive to planet mass; that sensitivity is amplified by a system's proximity to resonance. See \cite{Steffen15} for a more complete comparison of the sensitivities associated with TTVs and RVs. Thus, it is perhaps not surprising that the lowest density (i.e., lowest mass) sub-Saturns have more TTV than RV measurements. \citet{Lee15b} offer a parallel astrophysical explanation: TTV measurements are most easily made for systems in or near mean-motion resonances; such resonant systems formed by convergent inward migration; the planets comprising a resonant system therefore formed at larger orbital distances where disk gas was colder, less dense, and optically thinner; such gas cools more rapidly and is therefore accreted more readily onto rocky cores, forming especially low-density (''super-puffy'') planets (see also Section~\ref{ssec:formation-scenarios}).

\subsection{Follow-up Opportunities}
\subsubsection{TTVs}
\label{sssec:ttv}
Due to the proximity of \stellar{name}{b} \& c to the 2:1 mean-motion resonance, coherent gravitational interactions between the planets will result in large TTVs. \cite{Lithwick12} developed an analytic theory to describe TTVs near first order resonances, i.e.\ $j$+$1$:$j$ resonances where $j$ is an integer. We use this theory to estimate, within an order of magnitude, the amplitude of TTVs in this system. Following \cite{Lithwick12}, the normalized distance to resonance is given by 
\[
\Delta \equiv \frac{P^\prime}{P}\frac{j - 1}{j} - 1,
\]
where $P$ is the period of the inner planet and $P^{\prime}$ is the period of the outer planet. For the \stellar{name}{bc} pair, $j$~=~2 and $\Delta$~=~0.014. Near resonance, TTVs are oscillatory with a ``super period'' given by 
\[
P^{j} = \frac{P^\prime}{j\Delta}.
\]
\newcommand{\Zfree}{\ensuremath{Z_\mathrm{free}}\xspace}
For the \stellar{name}{bc} pair, $P^{j}$ is 1490~d or $\approx4$ years. 

Another important quantity that influences TTVs is \Zfree, a linear combination of the free complex eccentricities of the two planets.%
\footnote{For a more detailed discussion of \Zfree and how it relates to the forced and free eccentricities of both planets, see \cite{Lithwick12}}
When $|\Zfree| \ll |\Delta|$ or  $|\Zfree| \gg |\Delta|$ the amplitude of the TTV signal, $|V|$, is given by equations 14 and 15 in \cite{Lithwick12}:
\begin{eqnarray}
|V| \sim P \frac{\mu^{\prime}}{|\Delta|}\left(1 + \frac{|\Zfree|}{|\Delta|}\right) 
\end{eqnarray}
and
\begin{eqnarray}
|V^{\prime}| \sim P^{\prime} \frac{\mu}{|\Delta|}\left(1 + \frac{|\Zfree|}{|\Delta|}\right),
\end{eqnarray}
where $\mu = \Mp/\Mstar$. We consider a useful limiting case where $|\Zfree| \ll |\Delta|$. In this case, the respective TTV amplitudes of planets b and c are
\begin{eqnarray}
|V| \sim P \frac{\mu^{\prime}}{|\Delta|} \sim 2.6~\mathrm{hr}
\end{eqnarray}
and 
\begin{eqnarray}
|V^{\prime}| \sim P^{\prime} \frac{\mu}{|\Delta|} \sim 4.0~\mathrm{hr}.
\end{eqnarray}
Given that \stellar{name} is bright ($V$~=~11.3) and that the planets are large, TTVs of this magnitude are easily detectable from the ground. We emphasize that the TTVs can be significantly larger when $|\Zfree| \approx e > |\Delta| \approx 0.01$, and so the nominal TTV amplitude estimated above provides a rough lower limit to the TTV amplitude. Strictly speaking, however, there are some ``coincidental'' orbital configurations with $|\Zfree| \sim \Delta$ where the TTV amplitudes could be smaller than the above estimates. However, such configurations are rare (see \citealt{Lithwick12}). Given that $|\Zfree|$ depends on $e_{b}$, $e_{c}$, $\varpi_{b}$, and $\varpi_{c}$, observing and modeling the TTVs over an appreciable fraction of the 4~yr super period will place important constraints on the orbits of \stellar{name}{b} and c.

\subsubsection{Transmission Spectroscopy}
The fact that \stellar{name}{b} and c are large, low-density, and orbit a bright host star ($V$~=~11.3) makes them especially favorable targets for atmospheric characterization via transmission spectroscopy. Such observations could directly test our conclusions about the planets' bulk composition and formation history by measuring their atmospheres' elemental compositions and overall metal enrichments. For cloud-free, hydrogen-dominated atmospheres, we expect features in the transmission spectra to have amplitudes of $\sim 10 H \Rp / \Rstar^{2}$, where $H$ is the atmospheric scale height \citep{Miller-Ricci09}, which corresponds to $\sim$250~ppm and $\sim$400~ppm for planets b and c, respectively.

Features of this size should be detectable even with current instrumentation on the {\em Hubble Space Telescope}. We note that, given the high-altitude clouds or hazes frequently seen in exoplanet atmospheres, the spectral features of the \stellar{name} planets could be wholly muted at {\em HST}-accessible wavelengths. However, in just a few years, high-precision spectroscopy with {\em JWST} should be capable of detecting the strongest absorption features (e.g. by CO$_2$ at 4--5\,$\mu$m; \citealt{Morley15}). By measuring the thermal emission spectra,  MIRI should handily detect the planets' expected 10\,$\mu$m eclipse depths of $\sim$100~ppm (Greene et al. 2015, in press).  Future observations will measure the atmospheric makeup of these and other low-density sub-Saturns and so begin to elucidate the nature of these mysterious objects.

\section{Conclusions}
\label{sec:conclusions}
We have presented the discovery and characterization of two sub-Saturn-sized planets orbiting \stellar{name} detected by \ktwo in Campaign 2. We conducted follow-up adaptive optics imaging and spectroscopy of  \stellar{name} and found that it is a single, metal-rich (\fe~=~\stellar{fe}~dex) G3 star. We confirmed the two planets using Keck/HIRES by measuring the changes in the radial velocity of \stellar{name} due to its planets. Our RV measurements also constrain planet mass, density, and interior structure. \stellar{name}{b} has a size of \planet{b}{Rp}~\Re and a mass of \planet{b}{mass}~\Me. \stellar{name}{c} is larger and  more massive, having \Rp~=~\planet{b}{Rp}~\Re and \Mp~=~\planet{c}{mass}~\Me.

We combined the measured sizes and masses of \stellar{name}{b} and c with the interior structure models of \cite{Lopez13}, to constrain the likely distribution of planet mass between core and envelope. According to these models, $\sim$75\% of \stellar{name}{b}'s mass (\Mcoreb~\Me) is concentrated in its core. \stellar{name}{c} has a similar core mass, \Mcorec~\Me, but that only comprises $\sim$~50\% of its total mass. We explored the possible formation scenarios of \stellar{name}{b} \& c and hypothesize that the planets formed exterior to their current locations, and migrated inward as a resonant pair. We have difficulty explaining how \stellar{name}{c} nearly doubled in mass without undergoing runaway accretion to form a Jovian-mass planet. We encourage further follow-up of these planets using TTVs to constrain the orbits and dynamical state of the system and with transmission spectroscopy to measure atmospheric composition and structure. We also encourage further study of planets in the sub-Saturn size range. While the current sample of sub-Saturns around bright stars is quite small, upcoming \ktwo Campaigns along with future missions like {\em TESS} and {\em PLATO} should reveal many more.

\acknowledgements 
We thank Geoffrey Marcy, Konstantin Batygin, and Leslie Rogers for helpful discussions. We thank an anonymous referee for valuable comments. E.~A.~P.\ acknowledges support from a Hubble Fellowship grant HST-HF2-51365.001-A awarded by the Space Telescope Science Institute, which is operated by the Association of Universities for Research in Astronomy, Inc. for NASA under contract NAS 5-26555.  A.~W.~H.\ acknowledges support for our \ktwo team through a NASA Astrophysics Data Analysis Program grant.  A.~W.~H.\ and I.~J.~M.~C.\ acknowledge support from the \ktwo Guest Observer Program. E.~D.~L. received funding from the European Union Seventh Framework Programme (FP7/2007-2013) under grant agreement number 313014 (ETAEARTH). B.~J.~F.\ acknowledges support from a National Science Foundation Graduate Research Fellowship under grant No. 2014184874. This research used resources of the National Energy Research Scientific Computing Center, a DOE Office of Science User Facility  supported by the Office of Science of the U.S. Department of Energy  under Contract No. DE-AC02-05CH11231. This work made use of the SIMBAD database (operated at CDS, Strasbourg, France), NASA's Astrophysics Data System Bibliographic Services, and data products from the Two Micron All Sky Survey (2MASS), the APASS database, the SDSS-III project, and the Digitized Sky Survey. Some of the data presented in this paper were obtained from the Mikulski Archive for Space Telescopes (MAST). Support for MAST for non-HST data is provided by the NASA Office of Space Science via grant NNX09AF08G and by other grants and contracts. Some of the data presented herein were obtained at the W.~M.~Keck Observatory (which is operated as a scientific partnership among Caltech, UC, and NASA).The authors wish to recognize and acknowledge the very significant cultural role and reverence that the summit of Maunakea has always had within the indigenous Hawaiian community.  We are most fortunate to have the opportunity to conduct observations from this mountain.

{\it Facility:} \facility{\Kepler}, \facility{\ktwo}, \facility{Keck-I (HIRES)}, \facility{Keck-II (NIRC2)}

\bibliography{epic2037.bib}

\begin{thebibliography}{}
\expandafter\ifx\csname natexlab\endcsname\relax\def\natexlab#1{#1}\fi

\bibitem[{REV(????)}]{REVTEX41Control}
 ????

\bibitem[{08(1)}]{apsrev41Control}
08. 1

\bibitem[{{Albrecht} {et~al.}(2012){Albrecht}, {Winn}, {Johnson}, {Howard},
  {Marcy}, {Butler}, {Arriagada}, {Crane}, {Shectman}, {Thompson}, {Hirano},
  {Bakos}, \& {Hartman}}]{Albrecht12}
{Albrecht}, S., {Winn}, J.~N., {Johnson}, J.~A., {et~al.} 2012, \apj, 757, 18

\bibitem[{{Bakos} {et~al.}(2015){Bakos}, {Penev}, {Bayliss}, {Hartman}, {Zhou},
  {Brahm}, {Mancini}, {de Val-Borro}, {Bhatti}, {Jord{\'a}n}, {Rabus},
  {Espinoza}, {Csubry}, {Howard}, {Fulton}, {Buchhave}, {Ciceri}, {Henning},
  {Schmidt}, {Isaacson}, {Noyes}, {Marcy}, {Suc}, {Howe}, {Burrows},
  {L{\'a}z{\'a}r}, {Papp}, \& {S{\'a}ri}}]{Bakos15}
{Bakos}, G.~{\'A}., {Penev}, K., {Bayliss}, D., {et~al.} 2015, \apj, 813, 111

\bibitem[{{Barnes} \& {Greenberg}(2007)}]{Barnes}
{Barnes}, R., \& {Greenberg}, R. 2007, \apjl, 665, L67

\bibitem[{{Byrd} {et~al.}(1995){Byrd}, P.~{Lu}, \& J.}]{Byrd95}
{Byrd}, R.~H., P.~{Lu}, P., \& J., N. 1995, SIAM Journal on Scientific and
  Statistical Computing, 16, 1190

\bibitem[{{Cochran} {et~al.}(2011){Cochran}, {Fabrycky}, {Torres}, {Fressin},
  {D{\'e}sert}, {Ragozzine}, {Sasselov}, {Fortney}, {Rowe}, {Brugamyer},
  {Bryson}, {Carter}, {Ciardi}, {Howell}, {Steffen}, {Borucki}, {Koch}, {Winn},
  {Welsh}, {Uddin}, {Tenenbaum}, {Still}, {Seager}, {Quinn}, {Mullally},
  {Miller}, {Marcy}, {MacQueen}, {Lucas}, {Lissauer}, {Latham}, {Knutson},
  {Kinemuchi}, {Johnson}, {Jenkins}, {Isaacson}, {Howard}, {Horch}, {Holman},
  {Henze}, {Haas}, {Gilliland}, {Gautier}, {Ford}, {Fischer}, {Everett},
  {Endl}, {Demory}, {Deming}, {Charbonneau}, {Caldwell}, {Buchhave}, {Brown},
  \& {Batalha}}]{Cochran11}
{Cochran}, W.~D., {Fabrycky}, D.~C., {Torres}, G., {et~al.} 2011, \apjs, 197, 7

\bibitem[{{Coelho} {et~al.}(2005){Coelho}, {Barbuy}, {Mel{\'e}ndez},
  {Schiavon}, \& {Castilho}}]{Coelho05}
{Coelho}, P., {Barbuy}, B., {Mel{\'e}ndez}, J., {Schiavon}, R.~P., \&
  {Castilho}, B.~V. 2005, \aap, 443, 735

\bibitem[{{Crossfield} {et~al.}(2015){Crossfield}, {Petigura}, {Schlieder},
  {Howard}, {Fulton}, {Aller}, {Ciardi}, {L{\'e}pine}, {Barclay}, {de Pater},
  {de Kleer}, {Quintana}, {Christiansen}, {Schlafly}, {Kaltenegger}, {Crepp},
  {Henning}, {Obermeier}, {Deacon}, {Weiss}, {Isaacson}, {Hansen}, {Liu},
  {Greene}, {Howell}, {Barman}, \& {Mordasini}}]{crossfield:2015}
{Crossfield}, I.~J.~M., {Petigura}, E., {Schlieder}, J.~E., {et~al.} 2015,
  \apj, 804, 10

\bibitem[{{Cumming} {et~al.}(2008){Cumming}, {Butler}, {Marcy}, {Vogt},
  {Wright}, \& {Fischer}}]{Cumming08}
{Cumming}, A., {Butler}, R.~P., {Marcy}, G.~W., {et~al.} 2008, \pasp, 120, 531

\bibitem[{{Deck} \& {Batygin}(2015)}]{Deck15}
{Deck}, K.~M., \& {Batygin}, K. 2015, \apj, 810, 119

\bibitem[{{Delfosse} {et~al.}(1998){Delfosse}, {Forveille}, {Mayor}, {Perrier},
  {Naef}, \& {Queloz}}]{Delfosse98}
{Delfosse}, X., {Forveille}, T., {Mayor}, M., {et~al.} 1998, \aap, 338, L67

\bibitem[{{Dotter} {et~al.}(2008){Dotter}, {Chaboyer}, {Jevremovi{\'c}},
  {Kostov}, {Baron}, \& {Ferguson}}]{Dotter08}
{Dotter}, A., {Chaboyer}, B., {Jevremovi{\'c}}, D., {et~al.} 2008, \apjs, 178,
  89

\bibitem[{{Eastman} {et~al.}(2013){Eastman}, {Gaudi}, \& {Agol}}]{Eastman13}
{Eastman}, J., {Gaudi}, B.~S., \& {Agol}, E. 2013, \pasp, 125, 83

\bibitem[{{Foreman-Mackey} {et~al.}(2013{\natexlab{a}}){Foreman-Mackey},
  {Hogg}, {Lang}, \& {Goodman}}]{foreman-mackey:2012}
{Foreman-Mackey}, D., {Hogg}, D.~W., {Lang}, D., \& {Goodman}, J.
  2013{\natexlab{a}}, \pasp, 125, 306

\bibitem[{{Foreman-Mackey} {et~al.}(2013{\natexlab{b}}){Foreman-Mackey},
  {Hogg}, {Lang}, \& {Goodman}}]{Foreman-Mackey13}
---. 2013{\natexlab{b}}, \pasp, 125, 306

\bibitem[{{Fressin} {et~al.}(2013){Fressin}, {Torres}, {Charbonneau}, {Bryson},
  {Christiansen}, {Dressing}, {Jenkins}, {Walkowicz}, \& {Batalha}}]{Fressin13}
{Fressin}, F., {Torres}, G., {Charbonneau}, D., {et~al.} 2013, \apj, 766, 81

\bibitem[{{Gladman}(1993)}]{Gladman}
{Gladman}, B. 1993, \icarus, 106, 247

\bibitem[{{Han} {et~al.}(2014){Han}, {Wang}, {Wright}, {Feng}, {Zhao},
  {Fakhouri}, {Brown}, \& {Hancock}}]{Han14}
{Han}, E., {Wang}, S.~X., {Wright}, J.~T., {et~al.} 2014, \pasp, 126, 827

\bibitem[{{Howard} {et~al.}(2010){Howard}, {Johnson}, {Marcy}, {Fischer},
  {Wright}, {Bernat}, {Henry}, {Peek}, {Isaacson}, {Apps}, {Endl}, {Cochran},
  {Valenti}, {Anderson}, \& {Piskunov}}]{howard:2010b}
{Howard}, A.~W., {Johnson}, J.~A., {Marcy}, G.~W., {et~al.} 2010, \apj, 721,
  1467

\bibitem[{{Howard} {et~al.}(2012){Howard}, {Marcy}, {Bryson}, {Jenkins},
  {Rowe}, {Batalha}, {Borucki}, {Koch}, {Dunham}, {Gautier}, {Van Cleve},
  {Cochran}, {Latham}, {Lissauer}, {Torres}, {Brown}, {Gilliland}, {Buchhave},
  {Caldwell}, {Christensen-Dalsgaard}, {Ciardi}, {Fressin}, {Haas}, {Howell},
  {Kjeldsen}, {Seager}, {Rogers}, {Sasselov}, {Steffen}, {Basri},
  {Charbonneau}, {Christiansen}, {Clarke}, {Dupree}, {Fabrycky}, {Fischer},
  {Ford}, {Fortney}, {Tarter}, {Girouard}, {Holman}, {Johnson}, {Klaus},
  {Machalek}, {Moorhead}, {Morehead}, {Ragozzine}, {Tenenbaum}, {Twicken},
  {Quinn}, {Isaacson}, {Shporer}, {Lucas}, {Walkowicz}, {Welsh}, {Boss},
  {Devore}, {Gould}, {Smith}, {Morris}, {Prsa}, {Morton}, {Still}, {Thompson},
  {Mullally}, {Endl}, \& {MacQueen}}]{Howard12}
{Howard}, A.~W., {Marcy}, G.~W., {Bryson}, S.~T., {et~al.} 2012, \apjs, 201, 15

\bibitem[{{Howard} {et~al.}(2014){Howard}, {Marcy}, {Fischer}, {Isaacson},
  {Muirhead}, {Henry}, {Boyajian}, {von Braun}, {Becker}, {Wright}, \&
  {Johnson}}]{Howard14}
{Howard}, A.~W., {Marcy}, G.~W., {Fischer}, D.~A., {et~al.} 2014, \apj, 794, 51

\bibitem[{{Howell} {et~al.}(2014){Howell}, {Sobeck}, {Haas}, {Still},
  {Barclay}, {Mullally}, {Troeltzsch}, {Aigrain}, {Bryson}, {Caldwell},
  {Chaplin}, {Cochran}, {Huber}, {Marcy}, {Miglio}, {Najita}, {Smith},
  {Twicken}, \& {Fortney}}]{howell:2014}
{Howell}, S.~B., {Sobeck}, C., {Haas}, M., {et~al.} 2014, \pasp, 126, 398

\bibitem[{{Huber} {et~al.}(2013){Huber}, {Chaplin}, {Christensen-Dalsgaard},
  {Gilliland}, {Kjeldsen}, {Buchhave}, {Fischer}, {Lissauer}, {Rowe},
  {Sanchis-Ojeda}, {Basu}, {Handberg}, {Hekker}, {Howard}, {Isaacson},
  {Karoff}, {Latham}, {Lund}, {Lundkvist}, {Marcy}, {Miglio}, {Silva Aguirre},
  {Stello}, {Arentoft}, {Barclay}, {Bedding}, {Burke}, {Christiansen},
  {Elsworth}, {Haas}, {Kawaler}, {Metcalfe}, {Mullally}, \&
  {Thompson}}]{Huber13}
{Huber}, D., {Chaplin}, W.~J., {Christensen-Dalsgaard}, J., {et~al.} 2013,
  \apj, 767, 127

\bibitem[{{Inamdar} \& {Schlichting}(2015)}]{Inamdar15}
{Inamdar}, N.~K., \& {Schlichting}, H.~E. 2015, \mnras, 448, 1751

\bibitem[{{Isaacson} \& {Fischer}(2010)}]{Isaacson10}
{Isaacson}, H., \& {Fischer}, D. 2010, \apj, 725, 875

\bibitem[{Jones {et~al.}(2001)Jones, Oliphant, Peterson, {et~al.}}]{Jones01}
Jones, E., Oliphant, T., Peterson, P., {et~al.} 2001, {SciPy}: Open source
  scientific tools for {Python}, [Online; accessed 2015-10-21]

\bibitem[{{Kolbl} {et~al.}(2015){Kolbl}, {Marcy}, {Isaacson}, \&
  {Howard}}]{kolbl:2015}
{Kolbl}, R., {Marcy}, G.~W., {Isaacson}, H., \& {Howard}, A.~W. 2015, \aj, 149,
  18

\bibitem[{{Kreidberg}(2015)}]{Kreidberg15}
{Kreidberg}, L. 2015, ArXiv e-prints, arXiv:1507.08285

\bibitem[{{Lee} \& {Chiang}(2015{\natexlab{a}})}]{Lee15b}
{Lee}, E.~J., \& {Chiang}, E. 2015{\natexlab{a}}, ArXiv e-prints,
  arXiv:1510.08855

\bibitem[{{Lee} \& {Chiang}(2015{\natexlab{b}})}]{Lee15a}
---. 2015{\natexlab{b}}, \apj, 811, 41

\bibitem[{{Lee} {et~al.}(2014){Lee}, {Chiang}, \& {Ormel}}]{Lee14}
{Lee}, E.~J., {Chiang}, E., \& {Ormel}, C.~W. 2014, \apj, 797, 95

\bibitem[{{Liddle}(2004)}]{Liddle04}
{Liddle}, A.~R. 2004, \mnras, 351, L49

\bibitem[{{Lithwick} {et~al.}(2012){Lithwick}, {Xie}, \& {Wu}}]{Lithwick12}
{Lithwick}, Y., {Xie}, J., \& {Wu}, Y. 2012, \apj, 761, 122

\bibitem[{{Liu} {et~al.}(2015){Liu}, {Hori}, {Lin}, \& {Asphaug}}]{Liu15}
{Liu}, S.-F., {Hori}, Y., {Lin}, D.~N.~C., \& {Asphaug}, E. 2015, \apj, 812,
  164

\bibitem[{{Lopez} \& {Fortney}(2013)}]{Lopez13}
{Lopez}, E.~D., \& {Fortney}, J.~J. 2013, \apj, 776, 2

\bibitem[{{Lopez} \& {Fortney}(2014)}]{Lopez14}
---. 2014, \apj, 792, 1

\bibitem[{{Lucy} \& {Sweeney}(1971)}]{Lucy71}
{Lucy}, L.~B., \& {Sweeney}, M.~A. 1971, \aj, 76, 544

\bibitem[{{Marchal} \& {Bozis}(1982)}]{Marchal}
{Marchal}, C., \& {Bozis}, G. 1982, Celestial Mechanics, 26, 311

\bibitem[{{Marcy} \& {Butler}(1992)}]{Marcy92}
{Marcy}, G.~W., \& {Butler}, R.~P. 1992, \pasp, 104, 270

\bibitem[{{Marcy} {et~al.}(2001){Marcy}, {Butler}, {Fischer}, {Vogt},
  {Lissauer}, \& {Rivera}}]{Marcy01}
{Marcy}, G.~W., {Butler}, R.~P., {Fischer}, D., {et~al.} 2001, \apj, 556, 296

\bibitem[{{Marcy} {et~al.}(1998){Marcy}, {Butler}, {Vogt}, {Fischer}, \&
  {Lissauer}}]{Marcy98}
{Marcy}, G.~W., {Butler}, R.~P., {Vogt}, S.~S., {Fischer}, D., \& {Lissauer},
  J.~J. 1998, \apjl, 505, L147

\bibitem[{{Milani} \& {Nobili}(1983)}]{Milani}
{Milani}, A., \& {Nobili}, A.~M. 1983, Celestial Mechanics, 31, 213

\bibitem[{{Miller-Ricci} {et~al.}(2009){Miller-Ricci}, {Seager}, \&
  {Sasselov}}]{Miller-Ricci09}
{Miller-Ricci}, E., {Seager}, S., \& {Sasselov}, D. 2009, \apj, 690, 1056

\bibitem[{{Morley} {et~al.}(2015){Morley}, {Fortney}, {Marley}, {Zahnle},
  {Line}, {Kempton}, {Lewis}, \& {Cahoy}}]{Morley15}
{Morley}, C.~V., {Fortney}, J.~J., {Marley}, M.~S., {et~al.} 2015, ArXiv
  e-prints, arXiv:1511.01492

\bibitem[{{Morton}(2015)}]{Morton15}
{Morton}, T.~D. 2015, {isochrones: Stellar model grid package}, Astrophysics
  Source Code Library, ascl:1503.010

\bibitem[{{Murray} \& {Dermott}(2000)}]{Murray00}
{Murray}, C.~D., \& {Dermott}, S.~F. 2000, {Solar System Dynamics}

\bibitem[{{Parviainen} \& {Aigrain}(2015)}]{Parviainen15}
{Parviainen}, H., \& {Aigrain}, S. 2015, \mnras, 453, 3821

\bibitem[{{Petigura}(2015)}]{Petigura15thesis}
{Petigura}, E.~A. 2015, PhD thesis, University of California, Berkeley

\bibitem[{{Petigura} {et~al.}(2013){Petigura}, {Howard}, \&
  {Marcy}}]{Petigura13b}
{Petigura}, E.~A., {Howard}, A.~W., \& {Marcy}, G.~W. 2013, Proceedings of the
  National Academy of Science, 110, 19273

\bibitem[{{Petigura} {et~al.}(2015){Petigura}, {Schlieder}, {Crossfield},
  {Howard}, {Deck}, {Ciardi}, {Sinukoff}, {Allers}, {Best}, {Liu}, {Beichman},
  {Isaacson}, {Hansen}, \& {L{\'e}pine}}]{Petigura15}
{Petigura}, E.~A., {Schlieder}, J.~E., {Crossfield}, I.~J.~M., {et~al.} 2015,
  \apj, 811, 102

\bibitem[{{Pollack} {et~al.}(1996){Pollack}, {Hubickyj}, {Bodenheimer},
  {Lissauer}, {Podolak}, \& {Greenzweig}}]{Pollack96}
{Pollack}, J.~B., {Hubickyj}, O., {Bodenheimer}, P., {et~al.} 1996, Icarus,
  124, 62

\bibitem[{{Rivera} {et~al.}(2010){Rivera}, {Laughlin}, {Butler}, {Vogt},
  {Haghighipour}, \& {Meschiari}}]{Rivera10}
{Rivera}, E.~J., {Laughlin}, G., {Butler}, R.~P., {et~al.} 2010, \apj, 719, 890

\bibitem[{Schwarz(1978)}]{Schwartz78}
Schwarz, G. 1978, Annals of Statistics, 6, 461

\bibitem[{{Steffen}(2015)}]{Steffen15}
{Steffen}, J.~H. 2015, ArXiv e-prints, arXiv:1510.04750

\bibitem[{{Torres} {et~al.}(2012){Torres}, {Fischer}, {Sozzetti}, {Buchhave},
  {Winn}, {Holman}, \& {Carter}}]{Torres12}
{Torres}, G., {Fischer}, D.~A., {Sozzetti}, A., {et~al.} 2012, \apj, 757, 161

\bibitem[{{Valenti} {et~al.}(1995){Valenti}, {Butler}, \& {Marcy}}]{Valenti95}
{Valenti}, J.~A., {Butler}, R.~P., \& {Marcy}, G.~W. 1995, \pasp, 107, 966

\bibitem[{{Valenti} \& {Fischer}(2005)}]{Valenti05}
{Valenti}, J.~A., \& {Fischer}, D.~A. 2005, \apjs, 159, 141

\bibitem[{{Vogt} {et~al.}(1994){Vogt}, {Allen}, {Bigelow}, {Bresee}, {Brown},
  {Cantrall}, {Conrad}, {Couture}, {Delaney}, {Epps}, {Hilyard}, {Hilyard},
  {Horn}, {Jern}, {Kanto}, {Keane}, {Kibrick}, {Lewis}, {Osborne},
  {Pardeilhan}, {Pfister}, {Ricketts}, {Robinson}, {Stover}, {Tucker}, {Ward},
  \& {Wei}}]{Vogt94}
{Vogt}, S.~S., {Allen}, S.~L., {Bigelow}, B.~C., {et~al.} 1994, 2198, 362

\bibitem[{{Weiss} \& {Marcy}(2014)}]{weiss:2014}
{Weiss}, L.~M., \& {Marcy}, G.~W. 2014, \apjl, 783, L6

\bibitem[{{Winn} {et~al.}(2010){Winn}, {Johnson}, {Howard}, {Marcy}, {Bakos},
  {Hartman}, {Torres}, {Albrecht}, \& {Narita}}]{Winn10}
{Winn}, J.~N., {Johnson}, J.~A., {Howard}, A.~W., {et~al.} 2010, \apj, 718, 575

\bibitem[{{Wisdom} \& {Holman}(1991)}]{WisdomHolman}
{Wisdom}, J., \& {Holman}, M. 1991, \aj, 102, 1528

\bibitem[{{Wisdom} {et~al.}(1996){Wisdom}, {Holman}, \&
  {Touma}}]{SymplecticCorrector}
{Wisdom}, J., {Holman}, M., \& {Touma}, J. 1996, Fields Institute
  Communications, Vol.~10, p.~217, 10, 217

\bibitem[{{Zacharias} {et~al.}(2012){Zacharias}, {Finch}, {Girard}, {Henden},
  {Bartlett}, {Monet}, \& {Zacharias}}]{zacharias:2012}
{Zacharias}, N., {Finch}, C.~T., {Girard}, T.~M., {et~al.} 2012, VizieR Online
  Data Catalog, 1322, 0

\end{thebibliography}
\end{document}